\documentstyle[12pt,aasms]{article}

\def\la{\mathrel{\mathpalette\fun <}}
\def\ga{\mathrel{\mathpalette\fun >}}

\def\fun#1#2{\lower0.837ex\vbox{\baselineskip0ex\lineskip0.209ex
  \ialign{$\mathsurround=0ex#1\hfil##\hfil$\crcr#2\crcr\sim\crcr}}}

\def\msun{M_\odot}
\def\msunyr{M_\odot \ {\rm yr}^{-1}}
\def\sles{\lower2pt\hbox{$\buildrel {\scriptstyle <}
   \over {\scriptstyle\sim}$}}
 
\def\sgreat{\lower2pt\hbox{$\buildrel {\scriptstyle >}
   \over {\scriptstyle\sim}$}}

\def\la{\mathrel{\mathpalette\fun <}}
\def\ga{\mathrel{\mathpalette\fun >}}

\begin{document}

\title{ The Accretion Disk Limit Cycle Mechanism
   in the Black Hole
     X-ray   Binaries:
Toward an Understanding of the Systematic Effects}

\author{  John K. Cannizzo$^1$}
\affil{e-mail: cannizzo@lheavx.gsfc.nasa.gov}
\affil{Goddard Space Flight Center}
\affil{NASA/GSFC/Laboratory for High Energy Astrophysics, 
Code 662, Greenbelt, MD 20771}
\authoraddr{NASA/GSFC/Laboratory for High Energy Astrophysics, 
Code 662, Greenbelt, MD 20771}

\vskip 1truein
\ {\it }$^1$ Universities Space Research Association
\vskip 1truein
\centerline{ to appear in the Astrophysical Journal,
  1998, February 10, vol. 494}
\received{ 1997 July 21      }
\accepted{ 1997 September 17 }

\begin{abstract}
We examine in detail several aspects of 
the physics of accretion disks
of possible relevance to the outburst
mechanism of the black hole
X-ray transients.
We adopt
the one dimensional,
time dependent model described
in detail by Cannizzo, Chen, \& Livio
with parameters
    appropriate
for a system such as A0620-00.
We investigate
(1) the effect of the grid spacing, utilizing
   a logarithmic radial spacing
       $\Delta r\propto r$
in addition to 
     the spacing $\Delta r\propto \sqrt{r}$,
(2) the dependence of the local flow speed
of gas
within the hot part of the disk 
    on radius and time
during the  time of the
cooling wave propagation,
(3) the shape of the outburst light curve
       as a function  of the triggering location
       for the instability,
(4) the long term light curves of 
outbursts taken from trials in which complete
cycles of quiescence and outburst are followed,
both with and without including
    the effect
of     evaporation or removal of matter from 
the inner edge of the disk,
          and
(5) the strength of the self-irradiation
of the outer parts of the disk by
the X-rays from the inner disk.
Our primary findings in each of these areas
are that
(1) low resolution
runs taking $N\simeq20$ grid points
     using the logarithmic spacing
produce decay time scales
which are artificially
slow by factors of $\sim2-3$
and slower-than-exponential,
(2) the deviation from steady state within the 
outer part of the inner 
      hot disk appears
to be in accord with the discussion given in 
Vishniac \& Wheeler $-$ far from the transition
front the flow speed is $\sim\alpha c_s(h/r)$,
whereas at the interface between the transition
front and the cold disk
the flow speed is $\sim\alpha c_s $,
(3) the outburst triggering location must be
  $\ga100r_{\rm inner}$ for the rise
time of the resulting outburst to be as short
as is observed in the standard, bright systems,
(4) the long term light curves using the standard
model produce frequent
     outbursts which are triggered
near the inner disk edge and which have slow rise times,
and the long term light curves calculated assuming
evaporation of matter from the inner disk exhibit
outbursts  with longer 
         recurrence times and somewhat
(but not significantly)
      shorter rise times,
      and
(5) for a system with parameters
relevant to A0620-00, our ``standard'' system,
   irradiation is not a dynamically significant
effect $-$ in accord with recent results by 
van Paradijs.

\end{abstract}

\medskip
\medskip


{\it  Subject headings:}
accretion,  accretion disks $-$ binaries: close  $-$
           black hole physics  $-$ stars: individual (A0620-00)
   $-$  X-rays: bursts $-$ X-rays: stars

\section{ INTRODUCTION }

The black hole X-ray binaries (BHXBs)
are interacting binary systems
in which a Roche lobe filling
K$-$ or M$-$type secondary
transfers
matter through the inner Lagrangian
point into orbit around a black hole
accretor.
The systems are therefore similar in many
respects to the cataclysmic variables,
the difference being that the accretor
is not a white dwarf.
  (For a critical
  comparison of soft X-ray transients
  and dwarf novae,
  see van Paradijs \& Verbunt 1984.)
The BHXBs were discovered because
of large outbursts seen first in X-rays
which lasted weeks to months.
The systems were
subsequently
studied optically, and found
to be binary systems
with orbital periods of several hours 
                     to several days,
and with large mass functions
$f(M) \ga 3\msun$ implying black hole primaries
(van Paradijs \& McClintock 1995).

The accretion disk limit
cycle mechanism which was
originally put forth to account
for the dwarf nova outbursts
(Meyer \& Meyer-Hofmeister 1981;
for recent reviews see Cannizzo 1993a, 1998b;
Osaki 1996)
should work   in
   other interacting binary stars.
The model basically operates
due to an inherent
hysteretic
relation between the vertically
integrated viscous stress $\nu\Sigma$
and the surface density $\Sigma$
at a fixed radius in the disk.
The quantity $\nu$ is the kinematic
  viscosity coefficient 
  $\nu=2\alpha P/(3\Omega\rho)$,
and $\alpha$ is the Shakura \& Sunyaev (1973)
  parameter  characterizing 
  the magnitude of the viscous stress.
  When equilibrium solutions
  are plotted as $\log\nu\Sigma$ 
       (or $\log T_{\rm effective}$)
versus $\log\Sigma$,
one sees an $S-$curve relation.
The lower and upper branches of the $S$
which have positive slope are thermally
and viscously stable,
whereas the middle negatively
sloping portion of the $S$ is 
unstable.
   During quiescence,
  the   gas in the accretion disk
  is far from steady state:
  matter accumulates at large radii
in the disk and almost none reaches
  the central object.
   The gas is neutral, and the  
 temperature  and surface density  
     values   it possesses
  place it along the lower stable
  portion of the $S$ curve.
  As matter continues to pile up, 
  at some point
  a critical surface density 
  $\Sigma_{\rm max}$
corresponding to the local maximum in the
  $S$ curve is exceeded, and a heating
  instability is initiated which
  ultimately takes all or most of the disk
  to the hot state wherein the gas is ionized
  and resides along the upper stable branch of the
  $S$ curve. The rate of viscous evolution
is much greater than before, and the 
  $\Sigma(r)$   profile shifts to a near
  steady state configuration in which
$\Sigma(r)$ decreases with radius,
  in contrast to the quiescent configuration
  in which $\Sigma(r)$ increases with radius.
  After some material has accreted onto the
  central object and produced an outburst,
the surface density decreases until
  at some point the local value 
drops below another critical value
  $\Sigma_{\rm min}$ characterizing the
  lower bend in the $S$ curve.
  This always happens near the outer disk edge,
 and initiates
  a cooling wave  which propagates
  inward to smaller radii, 
  shutting off the flow and reverting 
  the disk material back into neutral gas.
   If this model is correct,
one cannot avoid having the
limit cycle 
instability operating  $-$ unless
the mass transfer rate 
    feeding into the outer edge
    exceeds a critical value
  (Smak 1983), or irradiation
        is able to keep the
entire disk      permanently
ionized
(Tuchman, Mineshige, \& Wheeler 1990;
van Paradijs 1996).
Not long after its introduction,
various workers
began to apply the model
to the soft X-ray transients $-$
a   class
which includes systems
with both neutron star and black hole
accretors (Cannizzo, Wheeler, \& Ghosh 1982, 1985;
Lin \& Taam 1984;
Huang \& Wheeler 1989,
   hereafter HW;
Mineshige \& Wheeler 1989, hereafter MW).

Cannizzo, Chen, \& Livio (1995, hereafter CCL) investigated
in detail the decay of the light curves
accompanying the outbursts
in the BHXBs.
They discovered that,
if a cooling wave is responsible
for the decay of the outbursts,
then
the observed exponential
decay of the soft X-ray fluxes
in the BHXBs
constrains the 
viscosity parameter
$\alpha$ to
have the form
$\alpha=\alpha_0(h/r)^{1.5}$,
where $\alpha_0\simeq50$
if the primary (i.e., black hole)
mass in the BHXBs is $\sim10\msun$.
This form was introduced 
by Meyer \& Meyer-Hofmeister (1983, 1984).
The value of $\sim50$ for the
proportionality constant
is mandated by the observed
$\sim30$ d $e-$folding decay times
for outbursts in the BHXBs.
The constraint is actually on
$M_1/\alpha_0$.
A recent compilation of inferred
black hole masses
for BHXBs by Bailyn et al.
(1996) seems to indicate
that the distribution is strongly
peaked at $\sim7\msun$,
whereas the value $\alpha_0\sim50$
was based on assuming $M_1\sim10\msun$.
If the work of Bailyn et al.
is borne out, then we would
need $\alpha_0\sim35$.
To maintain continuity with CCL,
however,
    we keep
$M_1=10\msun$ and $\alpha_0=50$.

        Vishniac \& Wheeler (1996,
   hereafter VW;
see also Vishniac 1997,
   hereafter V97)
present   an  important study
  which
provides a 
framework for understanding
    CCL's findings
based on considerations
of the deviation from steady state
flow conditions
within the hot, inner
part of the accretion disk.
Their key insight is 
that the flow within the
hot part of the accretion disk
which exists at small radii
during the decay from
maximum light after
an instability has been triggered
exhibits a deviation from steady
state conditions
as one approaches
the interface with the cold
portion of the disk.
Their approach to 
the problem represents
 a fundamental
    overthrow of the 
  old idea of expressing
the transition front
speeds in terms of purely
local conditions (Meyer 1984).
Part of our motivation
in this work is to
understand and to test
their model.

CCL's model for the limit cycle
instability had an advantage over
previous studies
in that their numerical resolution
was much finer than in other models,
and they could, through systematic testing,
place limits on the extent to which
certain aspects of the model could be trusted.
In this work we continue with this line of inquiry
by running a series of models
based on the numerical code described 
in previous works (Cannizzo 1993b, 1994;
   hereafter C93b and C94, respectively;  CCL)
to investigate various aspects
of the model
which were studied by CCL.
In particular, we
study 
(i) the effects of grid spacing,
(ii) the analytical model presented by VW,
(iii) the shape of the outburst versus instability triggering
     radius,
(iv) long term light curves covering quiescence and outburst,
         both with and without  incorporating evaporation 
   of gas from the inner edge of the disk,
(v) and the strength of self-irradiation
    of the outer disk by flux from the inner disk.
In the Discussion section
   we attempt to synthesize
  the lessons that were learned
in the previous sections
into (somewhat of) a coherent picture
of the operation of the limit
cycle instability in the BHXBs.

\section { MODEL CALCULATIONS }

For our standard model we
use that presented in CCL.
The  code
has been
described extensively
(e.g., C93b, C94, CCL):
 it solves
explicitly
for the time dependent evolution
of surface density $\Sigma(r,t)$
and midplane temperature $T_{\rm mid}(r,t)$
by simultaneously calculating
small corrections in $\Sigma$ and $T_{\rm mid}$
based on the equations of mass and energy
conservation.
The scalings parametrizing the steady state
                equilibrium disk
structure are given in C93b and CCL,
and are based on Cannizzo \& Wheeler (1984),
    Meyer \& Meyer-Hofmeister (1981),
and 
Pringle, Verbunt, \& Wade (1986).

C93b carried
out a series of tests
using the time dependent
 model to determine which
  terms are dominant
  and which are less
important
   in their effect
   on the long term light 
   curves.  
   They also looked at 
  the effect of varying the number of
  grid points.
     The results of the extensive
      testing done by C93b
  have  been generally confirmed
  in a recent work which  uses  a more
     advanced time dependent
    disk   instability   code
  that solves
   not only the vertically averaged
  viscous and thermal energy 
  equations   but also  the radial 
  Navier-Stokes equation and therefore
   includes radial pressure gradients
  and departures from Keplerian flow
   within the transition fronts
  (Ludwig \& Meyer 1997).

We assume a central object mass
$M_1=10\msun$,
a feeding rate into the outer disk
  $3\times10^{-11}\msunyr$
   (McClintock et al. 1983),
    a viscosity
parameter $\alpha=50(h/r)^{1.5}$,
an inner radius $r_{\rm inner}=10^7$ cm,
  and
an outer radius $r_{\rm outer}=10^{11}$ cm.
  We
    use a grid spacing $\Delta r \propto \sqrt{r}$
in sections 2.2, 2.3, and 2.4,
and a spacing $\Delta r\propto r$ in
sections 2.1 and 2.5.
      In the three sections
   for which $\Delta r \propto \sqrt{r}$,
we take $N=1000$ grid points
   where we look at the properties
       of individual outbursts,
        whereas
we take $N=500$ grid points
 for the sections which treat
complete cycles of outburst and quiescence.
  For section 2.1, we take 
$N=2^m(10)+1$, where $1\le m \le 5$,
and for section 2.5 we take $m=3$ (i.e., $N=81$).

\subsection{ Logarithmic Grid Spacing}

CCL utilized a grid spacing
$\Delta r \propto \sqrt{r}$.
  They discuss
earlier time dependent work
by 
MW,
Ichikawa, Mineshige, \& Kato (1994;
   hereafter IMK),
and
Kim et al. (1994, 1996;
  hereafter K94 and K96, respectively)
in which a logarithmic spacing
$\Delta r \propto   r $
is adopted, taking $N\sim20-40$ grid points.
Some of the discussion in CCL
concerning the 
number of grid points required
to get believable results
was speculative 
in the sense that 
    CCL did not attempt
to run models with the
logarithmic spacing
to substantiate
their criticisms,
   but rather they
   extrapolated
from what they had found
using their high resolution
trials.
It seems worthwhile to
look briefly at
this issue of the logarithmic 
grid spacing so as to quantify
the requirements for good results.
Figure 1 of K94
shows a trial light curve
using $r_{\rm outer}/r_{\rm inner} = 10^4$,
$M_1 =3.2\msun$, and 
$\alpha=\alpha_0(h/r)^{1.5}$.
To compare our results more directly
with K94, we have generalized
our code so that the logarithmic spacing
can be used (Cannizzo 1996b).

Figure 1 shows the results of a series of
decays for $N=21$, 41, 81, 161, 321, and 501
grid points using the logarithmic spacing.
For ease of viewing the last run was started
from different initial conditions so as to be offset
slightly from the first five.
Several aspects of the decay are apparent.
The lower resolution runs have
a decay which is too slow.
For the $N=21$ trial,
the decay constant $t_e\sim60-100$ d,
whereas in the limit of large $N$
we see $t_e \rightarrow \sim30$ d.
Note also that in the low resolution
runs
one sees glitches in the light curves
and $t_e$ curves
associated with the turn-off of individual
grid points (``turn-off'' meaning the initiation
of the cooling transition).
Furthermore,
$t_e$ increases with time
within a given run for $N$ small, 
whereas we know from CCL and VW
that for the parameters
used, namely $\alpha\propto(h/r)^{1.5}$,
the decay constant $t_e$ should
not vary significantly
   in time.
Therefore,
the detailed decay rate in
studies such as those mentioned earlier
which used $N\sim20$ grid points
with the logarithmic spacing
have decays which are 
spuriously slow by a factor $\sim2-3$,
and for which the functional
form for the decay is more
concave upwards
(i.e., $t_e$ increasing with time)
than it should otherwise  be.

\subsection {A Close Look at the Cooling Front }

CCL present numerous tests
of the limit cycle model using
$\alpha=\alpha_0(h/r)^n$
for the Shakura-Sunyaev $\alpha$ parameter.
They show the results of experiments
for different values of 
$\alpha_0$ and $n$ and find that
for runs with $n$ close to 1.5,
the decays of the outbursts
have a nearly exponential form,
as required by observations.
CCL attempt to understand
this result by casting the
speed of the cooling front
in terms of its width.
VW show that this viewpoint
is not valid:
the width of the cooling front
is an artifact of its speed,
and not the cause.
A key part of VW's insight
into the physics of the cooling front
and outflow from the hot,
viscously evolving part of the disk
is that,
far away from the front the local
flow speed    $v_r\sim\alpha c_s(h/r)$,
whereas in the front the flow speed
$v_r\sim\alpha c_s$.
VW show how this leads to a front speed
\begin{equation}
 v_f = \alpha_f c_f \left( c_f \over {r_f \Omega_f } \right)^q,
\end{equation}
where the subscript $f$ denotes values within the cooling front,
    $1/q = 1 + (1+n/2)(a_1/[1-nb_1/2])$,
$n$ is the exponent associated with the $\alpha$ scaling, and
$a_1$ and $b_1$ are logarithmic derivatives
characterizing the steady state scaling associated
with the ionized state of the disk,
$a_1 = d\log T/d\log\Sigma$ and 
$b_1 = d\log T/d\log\alpha$ (VW; CCL).
In other words, the scaling for
the midplane temperature associated
with the hot disk is
$T_{\rm mid} \propto \Sigma^{a_1} \alpha^{b_1} r^{-c_1}$.
For the usual Kramer's law opacity,
$a_1=c_1=3/7$ and $b_1=1/7$ (C93b; CCL; VW).
The midplane temperature
associated with the local
minimum in the $\log T_{\rm eff} - 
\log\Sigma$ steady state curves
$T_{\rm min}\propto \alpha^{-1.1/7}
\Omega^{-3/70}$ (VW).
Assuming $\alpha\propto
(h/r)^n$
and using the fact that the sound speed
in the front $c_f\propto \sqrt{T_{\rm min}}$
yields $d\log T_{\rm min}/d\log r =$
$[9/140-(1.1/7)(n/2)]/[1+(1.1/7)(n/2)]$.
(This last expression which is eqn. [34] of VW
corrects an
error in eqn. [4] of CCL, namely the
$r$
exponent
      should have positive rather than
negative sign.
In addition, in eqn. [4] of CCL
there should not be a ``K'' on the right hand side.)

There are several predictions
of VW's theory
which are amenable to testing
with our time dependent code.
We examine first the evolution of
$v_r/\alpha c_s$
in space and time. 
Figure 2 shows the variation
of the local values
of 
$v_r/\alpha c_s$
and $h/r$ with radius
for eight time steps
spaced 10 d apart during the
early decay of a model outburst.
The local flow speed is in fact $\sim \alpha c_s (h/r)$,
the usual viscous flow speed,
some distance
       upstream from the transition front.
VW speculate that this point occurs
only
   slightly interior to $r_f$,
whereas we find it happens at $\sim (1/2) r_f$.
VW's derivation
is predicated on the fractional disk radius
at which this point occurs remaining constant
during the decay, which it does.
    Also, 
the flow speed does not attain
$\sim\alpha c_s$ at the inner
edge of the front,
as posited by VW,
but rather at the outer edge.
At the inner front edge the flow has
only attained a speed
$\sim (1/6) \alpha c_s$.
(In looking carefully
  at Fig. 2 
   one might have imagined a scaling
with $\sqrt{h/r}$ in the value of
  $v_r/c_s$ at the inner edge of the front,
but an extension of the run
  used to generate Fig. 2
  in which the cooling front  
  is followed down to $\sim10^9$ cm
shows that the shoulder in the $v_r$
curves which lies at the inner edge
of the cooling front maintains a constant
level in terms of $v_r/c_s$.)
%
%
   VW note  that CCL's
results hinted 
        that $\sim (1/6) \alpha c_s$
at  the inner front    edge,
and the more  detailed model presented
by V97 uses this information
  as a refinement.
Figure 3
shows, for the eight time steps
indicated in Fig. 2,
   the values
of the radii  at which the local
flow speed goes to zero
and the local flow speed
equals the standard
viscous value $\alpha c_s (h/r)$, 
both given in units of 
     the radius associated with the
outermost grid point in the hot state
$r(i_{\rm hot})$.
     In
    the early stages 
    of the outburst
the ratio
associated with the $v_r=0$ radius is $\ga0.6$,
but later it approaches a
limiting value $\sim0.4$.
  The formalism presented
  in VW which assumes $v_r\sim\alpha c_s$
for the flow speed at the inner edge
  of the cooling front finds
the $v_r=0$ fractional radius
to be $\sim0.23$,
whereas the formalism in V97 
which is ``corrected''
by assuming $v_r\sim(1/6)\alpha c_s$
at the inner edge of the cooling front
finds the $v_r=0$ fractional radius
to be $\sim0.36$,
   in accord with our results.
   The ratio associated with
   the point at which viscous flow
is reached asymptotes
to $\sim0.55$,
     smaller than that mentioned
       in VW.
The deviations seen
in the first few time steps
 from the limiting values
      are an artifact
of the transitory
phase near outburst maximum
when the outwardly moving heating
front has stopped near the
outer disk edge
and begun to reflect inward
  as a cooling
wave. 
  This artifact is also seen
  in  the width of the  
     transition
   front, which is quite
  broad near the end
  of the  peak
  of the outburst
just as the decay is beginning
(see the upper panel in Fig. 5
of Cannizzo 1996c).
Last, V97 note that
the rate of mass outflow
  at the front is about three times
  the rate of accretion
 onto the central object.
  This is in agreement with Fig. 9
of CCL,
and also Fig. 2c of C94.
In summary,
    the   discrepancies we find with VW
are minor and it seems
they have captured
  the essential physics.

A further prediction of VW is contained
within their expression
for the cooling front speed.
Their formula is more complicated
than one might have guessed from the
discussion in CCL.
One sees for instance a dependence on $a_1$.
From equation (1) we have that
$v_f\propto c_f^{q+n+1} r_f^{(q+n)/2}$.
      For the standard model parameters
given in CCL and reiterated
earlier,
    the optimal 
$n$ value for which $v_f\propto r_f$ is not
1.5 but rather $1.656$.   
In Figure 4 we show the results of varying
$n$ in much smaller increments than was 
attempted in CCL.
Eleven curves are given corresponding to
$n=$ 1.50, 1.52, ..., 1.70.
Rather than guessing an initial
configuration
close to outburst maximum and then following
the decay, as was done in CCL,
here we initiate an outburst at a small radius
and wait until the disk has gone
nearly entirely to the high state
and then begun its decay.
In this manner the disk adjusts to its desired
state
as the outburst decay starts,
and we are not left with a long transient near
the beginning of the decay as in CCL.
In Figure 4 we see that,
contrary to the discussion given 
in CCL,
$n=1.5$
does not yield a
precisely exponential decay.
Our optimal value is about 
 $1.63$, which is
    close to the value 1.656 given
by VW's analytical formalism.
Note the small dynamic range
in $e-$folding times represented along the
$y-$axis.
The observational uncertainties and intrinsic
source fluctuations would probably not allow
one to distinguish among the range of $n$
values shown here,
so this result is not of direct
relevance observationally.
The motivation is solely to confirm VW.
It would be a worthwhile exercise
    to 
analyze the decays
of dwarf novae and X-ray novae
in a consistent fashion to see how
precise of a constraint
one may be able to place on $n$,
or, more broadly,
how essential is the functional form
$\alpha=\alpha_0(h/r)^n$.

For completeness
we show in Figure 5 the variation
of the width of the
front during the
rise and decay of an outburst.
The outburst begins at a small
radius, so we see first the outward
propagation of the
heating front, then the
inward movement of the cooling front.
As noted by CCL, the width
of the cooling front
can be approximated
by $w\simeq\sqrt{hr}$
or $w/r\simeq\sqrt{h/r}$.
VW show that both the
cooling front width and
speed of propagation
are set by the departure
from steady state discussed
earlier
and 
    are not causally linked.
It is interesting that
the fractional width
of the outward moving
heating
front is given
  phenomenologically
     by
$w/r\simeq(h/r)^{0.75}$.
Ludwig \& Meyer (1997)
find somewhat broader
  heating fronts.
  Their Figure 14 is interesting
  because it plots the heating
  and cooling front velocities
  versus radius on a common scale.
 VW put forth an analytical theory
for the period of decaying light
associated with the inward
moving cooling front;
no such theory currently exists
for the period of rising light
associated with the outward moving
heating front.
The sharp spike in $\Sigma$
accompanying the heating waves
is caused by the
high viscosity material
pushing into the region
of lower viscosity material
(Lin, Papaloizou, \& Faulkner 1985).
The fact that it is narrower than the
cooling front  must be an artifact
of this pushing,
but further work
will be needed to determine
a valid numerical parametrization
of the speed in order to lead
ultimately to a successful
theory of the heating front.

\subsection {  Outburst Light Curve Versus Triggering Location}

It has been known for some 
time that,
within the context
of the limit cycle model,
outbursts
triggered near the inner
edge of the disk tend to show 
slow rise times,
while outbursts 
triggered at somewhat
larger radii
generate faster rise times
(Smak 1984; Cannizzo, Wheeler, \& Polidan 1986, 
   hereafter  CWP).
The reason for the difference in rise
times is detailed in CWP.
For most of the
bright BHXB outbursts,
one sees fast rise times
(of order a few days)
in outbursts lasting
$\sim100-200$ d.
It is therefore of interest
to find out what is the
minimum critical
radius such that a fast
rise will be obtained.
Figure 6 shows
 a  series of outbursts
generated by starting
with a smooth $\Sigma(r)$
distribution resembling
that expected at the end
of the quiescent (i.e., 
accumulation) phase,
and adding to this 
profile a small spike
whose maximum exceeds
the local value
of $\Sigma_{\rm max}$.
This is carried out by superposing
two functions,
as shown in the top panel of Fig. 6.
The  functions are
\begin{equation}
S_{\rm broad} = 0.7 \Sigma_{\rm max} 
\exp\left(-(\log(r(\rm cm)) - \log(r_{\rm mid}(\rm cm)))^2\over 2\right)
\end{equation}
and
\begin{equation}
S_{\rm narrow } = 1.1 \Sigma_{\rm max} 
\exp \left(-(r- r_{\rm center})^2 \over {2w^2}\right),
\end{equation}
where $r_{\rm mid}=10^9$ cm is the geometric
average of the inner and outer radii,
$r_{\rm center}$
is the central radius of the narrow Gaussian
used to trigger the outburst,
$w=r_{\rm center}/3$,
and $\Sigma_{\rm max}$
is the local maximum in $\Sigma$ from vertical structure
computations (Cannizzo \& Wheeler 1984).
For a given trial we prescribe
the initial surface density distribution
by taking $\Sigma(r) =$
$ \max( S_{\rm broad}, S_{\rm narrow})$.
Five trials were run,
corresponding to $\log r_{\rm center}$(cm)$=$
8, 9, 10, 10.25, and 10.5.
(For ease of viewing we do not show
the 10.25 and 10.5 curves
in the top panel of Fig. 6.)
Note the stark
difference in the rise time properties
between the first two
and last three curves.
The rise is slow  for the $\log r_{\rm center}$(cm)
$=$ 8 and 9 curves, whereas
it is fast for the $\log r_{\rm center}$(cm)
$=$ 10, 10.25, and 10.5 curves
(Cannizzo 1996b).
The instability is triggered at $t=0$
for all five trials,
but due to the finite travel time
of the heating front,
the effect of increased viscosity
and mass flow rate is not felt
at the inner disk until $\sim30-50$
d later for the three trials
with  triggering
at large radii,
whereas the effect is felt much sooner
for the trials triggered
at small radii.
The difference in the
heating wave properties is shown 
in Figure 7.
The two panels show the 
evolution of surface density,
in  5 d steps,
for the $\log r$(cm)$=$ 9 and 10 runs.
These two span the transition region
of parameter
space between slow and fast rise.
For the outbursts triggered at large radii,
the amplitude of the outburst
is larger because more mass is contained
within the wave of surface density
enhancement associated
with the heating front.
The slow rise times associated with the
$\log r$(cm)$=$ 8 and 9
curves
are caused
by the fact that,
to enhance appreciably the rate
of mass flow at the inner disk edge,
the outwardly traveling heating front
must access the bulk of the matter
which is stored in the disk
at  large radii,
between the Lubow \& Shu (1975) radius
(corresponding to that point in the
disk where the specific
angular momentum associated
with the gas leaving the
L1 point at the surface of the
secondary star
equals the Keplerian value
in the disk)
and the outer disk edge,
and this matter,
after having made the transition
from neutral to ionized gas,
then has to viscously evolve
(both to smaller and larger radii).
The physics associated with this flow is discussed
in CWP and C93a.

\subsection{ Long Term Light Curves  }


As noted by many authors,
the standard limit cycle model
fails 
    to account 
for observations of systems
in quiescence (see Lasota 1996a, 1996b).
One observes a significant
EUV and X-ray flux which we
infer to be the result of accretion during quiescence,
and yet if we are to take the theory
literally it predicts a quiescent
accretion rate which is negligibly small.
The small accretion rate
is mandated by the requirement that
   $\Sigma < \Sigma_{\rm max}$
and $T_{\rm eff} < T_{\rm eff}(\Sigma_{\rm max})$
for all radii,
in particular $r\sim r_{\rm  inner}$.
McClintock, Horne, \& Remillard (1995)
found evidence for low
level accretion
$\sim10^{10}-10^{11}$ g s$^{-1}$
(if the efficiency of accretion $\epsilon\sim0.1$)
in the quiescent state
of A0620-00,
but this is still much greater
than would be allowed 
in the ``pure'' version 
of the standard accretion
disk model in which
the inner disk edge
extends all the
way to the last stable
orbit
(Lasota, Narayan, \& Yi 1996).
The critical accretion rate
  associated with $\Sigma_{\rm min}$
is  ${\dot M}(\Sigma_{\rm min})\sim10^6$ g s$^{-1}$
$(r_{\rm inner}/10^7 {\rm \ cm})^{2.6}$
for BHXBs (CCL).
  There is nothing in the model,
     however,
which requires $r_{\rm inner}=6GM_1/c^2$.
  Time dependent computations
of the accretion disk limit cycle instability
have shown
that the model still  works
well if the inner edge is truncated
(e.g., Angelini \& Verbunt 1989;
C93b).
    There may exist some
physical mechanism
extrinsic to the standard
limit cycle model which serves to evaporate
or take material from the inner disk
and transfer it 
more or less directly
     onto the accretor
(e.g., Meyer 1990;  Meyer \& Meyer-Hofmeister 1994;
   Liu, Meyer, \& Meyer-Hofmeister 1995).
Such a mechanism may or may not
fit in  with 
the general paradigm of accretion
onto black holes
(Chakrabarti 1996a, 1996b; Narayan, McClintock, \& Yi 1996).
      Workers 
are beginning 
   to consider the effect
such a process
would have  on the limit cycle
model in terms of
how the altered surface
density distribution
  in quiescence $-$
  in particular, the lack of an inner disk $-$
would change the properties
of the outburst
   (Lasota et al. 1996;
   Hameury et al. 1997,
    hereafter HLMN).

When we run complete cycles using the
$\alpha=50  (h/r)^{1.5}$
scaling, we find frequent
    outbursts
which are triggered at small radii
and which have slow rise times,
in contrast to the observations.
If the inner disk were missing due to 
a slow removal of matter onto the accretor,
then the outburst would have to be triggered further out.
   This may not be a unique
solution to the problem,
and at least three  obvious alternatives
are apparent. (The problem 
of the quiescent X-ray  flux
would only be addressed by the third option.)

(1) The constraint imposed by observed exponential
decays $\alpha=50(h/r)^{1.5}$
is actually only applicable to the matter
which resides in the ``hot state'' part
of the disk.
Once the matter has made the transition
to the low state,
it does not participate
in the  flow dynamics.
Therefore it is quite probable
that $\alpha=50(h/r)^{1.5}$
is a scaling which applies
only to ionized gas
at $T> 10^4$ K,
and for neutral gas at $T<10^4$ K
there may be another scaling which drops $\alpha$
abruptly  to a value less 
than $50(h/r)^{1.5}$.
In this case,
the material would
evolve less in quiescence,
and there would
be 
 less tendency 
for the matter to spread
and trigger an inside-out outburst
with its attendant slow rise.

(2)  We have not yet activated 
 and tested  the section
of our code which provides for a variable
outer disk radius (see Ichikawa \& Osaki 1992).
The qualitative effect this has on the
evolution  is to sweep up
matter at large radii which
  shrinks 
   the outer disk in quiescence.
  The contraction is caused by  
    the accretion of
low angular momentum material.
This has the effect of enhancing the surface density
near $r_{\rm outer}$ and leaving
the disk more prone to outbursts triggered
at  large radii (see Ichikawa \& Osaki 1994).

(3) There may exist a mechanism
   which prevents the cooling wave from propagating
  once it reaches some radius $r_{\rm crit}$.
  Since ${\dot M}_{\rm min}(r)={\dot M}(\Sigma_{\rm min})$
   which provides a measure of the quiescent
   mass flow rate at a radius $r$
     scales as
  $\sim r^{2.6}$,
  if $r_{\rm crit}$
   were at $\sim10^9-10^{10}$ cm,
for   example,
  this would be sufficient to leave a small,
hot inner disk in quiescence that could give rise
  to the quiescent X-rays
at a level $\sim10^{10}-10^{11}$ g s$^{-1}$
      seen in A0620-00
using a standard accretion efficiency $\epsilon\sim0.1$.

Having made these concessions,
we must return to our original point
that the EUV/soft X-ray observations seem
to demand some evaporation
of the inner disk
(unless the third option just
discussed turns out to be true).
  Although we are uncertain of
  the physics which leads
to evaporation,
we should include
at   least an ad hoc  prescription
for removal of matter from the inner disk
in any complete time dependent   model.
  We therefore
  evacuate 
  matter from the
  inner disk
through the prescription
$\partial \Sigma(r)/\partial t = -\Sigma_0/(1+[r/r_0]^2)^2$,
where $\Sigma_0=10^2(M_1/\msun)^{-1}$ g cm$^{-2}$ s$^{-1}$
and 
$r_0 = 100 r_g = 2.95\times 10^7(M_1/\msun)$ cm
(R. Narayan 1996, private communication).
This
    gives an integrated mass removal
$ {\dot M}_{\rm evap} =$
$-\int (\partial\Sigma/\partial t) 2\pi r dr$
$ = \pi \Sigma_0 r_0^2/           (1+[r_{\rm inner}/r_0]^2)$
$=2.7\times 10^{17}$ g s$^{-1}$ $ (1+[r_{\rm inner}/r_0]^2)^{-1} (M_1/\msun)$.
We must 
have some matter 
  at each grid point
for our code to work properly,
   therefore 
     when $\Sigma$ drops below 1 g cm$^{-2}$
at a given radius we set it equal to 1 g cm$^{-2}$.

Figure 8 shows 
a series of light  curves
with and without evaporation.
We use the expression
given in the previous
paragraph for $ {\dot M}_{\rm evap}$,
with the normalization
multiplied by $10^{-x}$.
For the four panels shown,
$x=\infty$, 4, 3, and 2.
Hence in the top panel
we present our standard model without
   evaporation.
Outbursts are triggered
frequently $-$ about every $5$ yr $-$
   at small
radii in the disk.
    Most of the outbursts
are fairly low amplitude,
   but since the 1975 outburst
  of A0620-00 made it a sustained
   $\sim50$ Crab source in $3-6$ keV
X-rays as seen by {\it Ariel 5 }
(Elvis et al. 1975),
the fainter outbursts produced in
our ``standard model'' would have been seen
over the past $\sim20$ yrs,
had they occurred.
About every 50 yr
    we get
a major outburst
in the models.
The introduction
of evaporation
    removes    the inner parts of
the disk,
  eliminates the 
minor outbursts,
and increases
      the recurrence
times for the major outbursts.
Figure 9
shows the
mass of the accretion
disk accompanying the
runs shown in Fig. 8.
During a major outburst,
     a
fraction $\sim0.2$
of the disk
is accreted,
whereas during a minor outburst
$\la0.01$ of the disk
is accreted.

Figure 10 shows expanded
views of one major outburst
taken from each of the four panels
in Fig. 8.
The quiescent
values of the accretion
rate in third panel
of Fig. 10 show the
effective
rate of the evaporation or mass
loss from the disk.
In the models
of Narayan et al. (1996)
and Esin, McClintock, \& Narayan (1997),
      for instance,
the efficiency of accretion
is low in the quiescent
state $\epsilon \sim10^{-4}$,
while in outburst one still has $\epsilon\sim0.1$.
   Therefore
        one would see
a greater dynamic range
in soft X-ray flux between
quiescence and outburst
than would be inferred
directly from
the rates of accretion given in Fig. 10.
Also, our condition that $\Sigma(r)\ge1$
g cm$^{-2}$
   reduces
the effective
rate of evaporation
   increasingly  for smaller
    $x$ values
(i.e., higher  evaporation rate):
the effective evaporation rates
in going from $x=4$ to $x=2$
increase by a factor $\sim5$
for each of the two factor of 10 increases
in $10^{-x}$.

In Fig. 10 we see that,
as expected,
the fact that triggering
of outbursts cannot occur
at small radii
due to the truncation of the inner
disk
forces the rise times to shorten.
Even for the highest
values of ${\dot M}_{\rm evap}$,
however, the rise times
are $\sim1$ month $-$
     longer than the
value of  several days
 commonly seen.
For low values of the
evaporation, only the quiescent
surface density profile is affected.
The reason the outburst $\Sigma(r)$
profile is not  strongly
changed is that
during outburst
the surface density near the
inner edge is so large
that the local mass flow
greatly exceeds the local evaporation rate.
In looking at the decays
of the outbursts in the last panel of Fig. 10,
however, we see that
for a high  evaporation rate,
     the outburst itself is 
also affected:
the  decay is cut off
so that the exponential decay
is truncated.
This occurs because,
at some point during the inward
movement of the cooling front,
there is no longer any
disk left
into which the
cooling wave may propagate.
For instance,
for $x=2$  we only observe about 2 decades
of exponential decay in the value of ${\dot M}$
at the inner edge, which presumably powers the
soft X-ray flux we observe.
In systems like A0620-00, however,
we observe about 3 decades of exponential decay
(Tanaka \& Shibazaki 1996).
The model with $x=2$ corresponds to
${\dot M}_{\rm evap}\sim2\times 10^{14}$ g s$^{-1}$
with our standard assumed secondary star
       mass transfer rate value
        ${\dot M}_T \sim 2\times 10^{15}$ 
g s$^{-1}$.
The  rise time is $\sim1$ month.
The rise time could be shortened
presumably by increasing ${\dot M}_{\rm evap}$
further (i.e., decreasing $x$),
but that would
     reduce the dynamic range of the
soft X-ray flux accompanying the
exponential decay
to an even more unacceptably small value.

Figure 11 shows the evolution of 
 $\Sigma$ and $T_{\rm midplane}$
covering an outburst in the third
panel of Figure 8.
There are 10 time steps shown,
with the first time step
corresponding to
just before the triggering
  of the outburst.
The time increment $\delta t=0.1$ yr.
The starting time step shows the truncated
nature of the disk due to evaporation.
The outburst triggers at $\log r$(cm)$\sim9.6$.
In the early stages of the outburst
evaporation does not have any effect
on the outburst, as the $\Sigma(r)$ values
are quite large for $r$ small.
In looking at time steps 8, 9, and 10,
however,
we see that in the late stages of
the outburst 
when $\Sigma(r)$ becomes smaller,
    the evaporation
swallows up the last of the inner,
hot disk before the cooling front
can propagate all the way to the inner edge.
In time step 8 for instance the hot disk
exists only in a ring from about
8.5 to 9.5 in terms of $\log r$(cm).
For time steps 9 and 10,
      the entire disk
is back into the low state.
This leads to the truncation of the full
exponential decays evident in Figure 10.
Figure 12 reveals the evolution
  in quiescence by showing four time steps
which follow the last profile
      in Figure 11, here taking
$\delta t=10$ yr.
One can see evaporation eating into the
quiescent $\Sigma(r)$ distribution,
even as matter in the inner disk evolves viscously,
and matter at the outer edge continues to arrive
from the secondary star.

In summary,
 evaporation appears not quite capable, 
 at least  for a system with parameters
relevant for a system
like A0620-00,
of producing outbursts which occur at
large enough radii to give
     rise times
that are as fast as observed,
     and which simultaneously 
   have $\sim2-3$ decades of exponential decay.
 Since the models are so close to working,  
   one could argue that, given
    the crude nature of the models,
  we could consider  this a qualified
 success.
On the other hand, evaporation unquestionably
does help
     in the standard model
     as regards the recurrence times for outbursts.
   The effect or necessity of evaporation
     in the models may
       potentially  be different
for other model parameters,
for instance with systems for
larger orbital periods and higher
${\dot M}_T$ values.
HLMN
     find, for example,
that evaporation is  needed to
obtain light curves in various energy bands
that can account for observations of the
start of an outburst seen in GRO J1655-40,
a system with 
a much longer orbital period
   $P_{\rm orbital}=2.6$ d
(Bailyn et al. 1995)
and higher
    mass transfer rate 
${\dot M}_T=2\times 10^{17}$ g s$^{-1}$
than we have assumed 
in this work for our standard
model
($P_{\rm orbital}=7.75$ h giving
 $r_{\rm outer}=1.5\times10^{11}$ cm
and mass transfer rate
  ${\dot M}_T=1.89\times 10^{15}$ g s$^{-1}$)
     taken to be
    representative of A0620-00.
         HLMN
     do not show
light curves
covering complete cycles
of quiescence and outburst, 
    so it is difficult
to assess the full generality of their
conclusions.
In particular,
  it may prove difficult
   for any disk instability model
  to reproduce the prompt decay
  in $V$ which was seen in 
    {\it Hubble Space Telescope } observations
   that occurred even as the
 $2-10$ keV  X-ray flux was increasing
  for the 1996 outburst
   in GRO J1655-40
(Hynes et al. 1996).
   The fact that the two 
  microquasars
  do not show the classical fast-rise,
  exponential decay outburst may be
               associated with   
   the fact    that the BHs
  in these systems
   seem to be maximally rotating
   (Zhang, Cui, \& Chen 1997).
Our results do not support
the strength of  HLMN's contentions
about the necessity of evaporation.
Preliminary work
using our canonical SS Cygni  model
(C93b; Cannizzo 1996c, 1998c)
also finds evaporation
to be ineffectual in bringing about 
fast rise outbursts for models
which are intrinsically prone to
slow rise outbursts.
Cannizzo  (1998c) finds that,
even in
the limit where
      ${\dot M}_{\rm evap} \rightarrow {\dot M}_T$,
the outburst rise times as taken from
light curves computed using ${\dot M}_{\rm inner}$
   are still slower than the
fast-rise outburst  in SS Cygni
      observed by Mauche (1996)
in the EUV.
One could not of course
have ${\dot M}_{\rm evap} = {\dot M}_T$
or the entire disk would evaporate.

\subsection {Strength of the Irradiation}

Tuchman et al. (1990)
     present
a formalism
for incorporating
the effects
of   irradiation
on the equilibrium
structure
of the disk
determined by the vertical
structure
calculations.
We have included
their formalism 
into our time dependent model
to calculate
the local
irradiation temperature
and its effect
in altering
the S-curve.
In this preliminary work
we only activate the
portion of the code
which calculates
the irradiation
temperature.
We leave for future
work a computation
which includes
its  feedback
into the structure of the disk.

Shakura \& Sunyaev (1973)
give the local irradiation
flux as
\begin{equation}
F_{\rm irr} =  C {L_X \over {2\pi r^2}},
\end{equation}
where 
$C=A d(h/r)/d\ln r$,
$n=1$ for neutron star
accretors
and 
$n=2$ for black hole
accretors.
The parameter $A$ is the X-ray albedo.
The values of
$n$ advocated by Shakura \& Sunyaev
are based on
the supposed
  preferred 
   directions for the  emission:
isotropically
from the neutron star surface
  ($n=1$),
and primarily vertically
from the disk surface
for black hole primaries
    ($n=2$).
One cannot claim to  understand
the emission
geometry at small
radii in the BHXBs 
$-$ for instance 
there may exist a
scattering medium above the disk
which influences the
value  of   $n$ $-$
therefore we shows results
for both $n=1$ and 2
as being representative
of possible values.

Figures 13 and 14
show the values of 
ratios of the
locally  defined
irradiation temperature
$T_{\rm irr}=(F_{\rm irr}/\sigma)^{1/4}$
to the local (viscous)  effective
temperatures 
for 600 d of evolution
with $\delta t=6$ d
spanning the rise and
decay of an outburst
in the standard model.
We show $n=1$ and $n=2$,
and assume $A=1$.
   As in section 2.1,
we adopt a logarithmic grid spacing
so as to resolve better the
inner portions of the disk.
We take $N=81$ grid points.
For the case $n=2$
which is thought
to be more representative
of the BHXBs,
the local irradiation
temperature
is never more than about $0.3-0.4$
of the local effective
temperature
determined from
equating 
viscous heating
and radiative cooling.
This occurs near the time of maximum
  light in the outburst.
Van Paradijs (1996)
presents 
revised criteria
for the 
allowability 
of the limit cycle
mechanism in the  soft X-ray transients
including neutron star
and black hole systems.
He concludes
that 
irradiation
is more effective 
for neutron star accretors
than for BHXBs
at preventing
the limit cycle
from operating
by virtue of keeping
the entire disk in the ionized state.
In his Fig. 2
he indicates
A0620-00 as lying a factor
$\ga10$ below the dividing line
between steady and unsteady behavior.
Therefore our finding
of relatively weak irradiation
for a model meant to 
represent A0620-00 is consistent
with the fact that A0620-00
has exhibited two outbursts
(in 1917 and 1975)
and is  seemingly not
prone to irradiation induced effects.
It is also consistent with HW and MW
who found, by comparing
theory with observation
for A0620-00,
that irradiation
does not contribute
overwhelmingly to the optical flux.

\section {DISCUSSION }

We have investigated
a variety of
phenomena associated
with the limit cycle model
as it may apply to the BHXBs.
We find the following:

(i) The grid spacing in which
$\Delta r \propto r$ does not appear to
be as convenient
as the spacing $\Delta  r\propto \sqrt{r}$
because the former overresolves
the innermost part of the disk,
resulting in long CPU times
for execution.
Low resolution runs with $N\simeq20-40$
grid points produce light curves
which show artificial  ``steps''
in the decay caused by the turn-off
of individual grid points.
In addition, the locally defined $e-$folding
decays times are too slow by $\sim2-3$,
  and the decay character is artificially
  slower-than-exponential.
As $N$ is made to increase,
the steps becomes smaller and disappear,
and the decay time scales
asymptote to a common value,
as expected in a numerical scheme
which converges properly.
We have only investigated properties
  associated with the
  decay. There may be other spurious systematic
   effects associated, for instance,
with the long term light curves
(i.e., the triggering radii
  and outburst rise times),
   or with the production
     of reflares or spikes
    in the light curve
  (e.g. K96)
   which would be smoothed
   out in a much higher resolution model.

(ii) The light curves
for outbursts have slow rise
times when triggered
at radii $\la 10^9$ cm
in the disk,
and fast rise times
when triggered at radii $\ga 10^{10}$ cm.
There is an abrupt change in the character
of the rise at some radius intermediate
between these two values.
In this context, it is somewhat of a misnomer
to refer to the two cases as ``inside-out''
and ``outside-in'' since, in both instances,
the triggering radius is much less than
the outer disk radius.
The outburst light curves of interest
in this work are typified by A0620-00
and show not only an exponential decay with
$\sim30-40$ day $e-$folding time scale,
but also a fast rise (i.e., several days).
These experiments in which we artificially
trigger the outbursts at a specific  radius
show that, to produce outbursts with rise times
as fast as those observed,
the triggering radius must be at least
as large as some value between about
$10^9$ and $10^{10}$ cm, for A0620-00 parameters.

(iii) The model of VW for the decay
of the outburst, based on considerations
involving a deviation from steady-state flow
in the inner part of the disk,
is substantially borne out in detailed testing.
Their basic premise is that the local flow speed
``accelerates'' from  the local value of  $\alpha c_s(h/r)$
                to    the local value of  $\alpha c_s$ within
the hot part of the disk.
VW conceive of the point at which  $v_r\sim \alpha c_s$
is attained as being at the inner edge of the cooling transition
region, whereas we find it to be at the outer edge.
This does not impact on VW's derivation.
More importantly, we find that the value of the radius
at which $v_r$ changes sign, expressed in units of 
$r(i_{\rm hot})$,  asymptotes to $\sim0.4$,
    whereas the fractional radius at which
   the local flow rate equals the standard
        viscous flow rate
  asymptotes  to $\sim0.55$.

  The work of CCL had as its underlying
  premise the assumption that the decay
  of the soft X-ray flux in the outbursts
  seen in the BHXBs is due to a cooling wave.
  With that as a working hypothesis,
  CCL showed that $\alpha_{\rm hot}\sim50(h/r)^{1.5}$
  is needed, and the $e-$folding decay time
  for $dM/dt$ is 
    $  \sim0.4(GM_1/\alpha_0)c_s^{-3}$
where $c_s\sim16$ km s$^{-1}$.
An alternative starting point,
  advocated recently by King \& Ritter (1998),
is that the decays in the BHXB outbursts
are due to the gradual loss of disk material
into the BH.
  The cooling   wave is unable to start
because of strong irradiation,
  and so the hot, viscous disk can only
  lose material from its inner edge.
Cannizzo, Lee, \&  Goodman (1990)
showed that, for $\alpha$ constant,
$dM/dt$ decreases in a purely viscous disk
as $\sim t^{-1.3}$.
To get an exponential decay as observed,
the kinematic viscosity coefficient
$\nu\propto\alpha T_{\rm midplane}$ must be constant,
implying  either an isothermal disk
with $\alpha$ constant,
or else $\alpha_{\rm    }\propto 1/T_{\rm midplane}$.
Neither option seems physically reasonable.
It is interesting that for both the cooling wave
model and  the viscous decay model,
     a constant $\alpha_{\rm hot}$ value does not
      work (Cannizzo 1998b).
Mineshige, Yamasaki, \& Ishizaka (1993)
pointed out that purely
 viscous decays lead to power law
  light curves and that a cooling
wave which extracts
mass and angular momentum
 from the hot, viscous disk
  can produce the required
  exponential decay.
   They present a toy model
  in which $\partial\Sigma/\partial t$
  $  \propto -\Sigma/t_0$ is put
  in by hand, rather than
  determining a form for  $\alpha_{\rm hot}$
  that will lead to the exponential process
   self-consistently using a full
time dependent model as in CCL.
King \& Ritter (1998)
note the discrepancy
between the X-ray to optical ratio
in the  computations of HW,
and the observed value for A0620-00
in the 1975 outburst.
The main problem is that HW's
calculated peak ${\dot M}(r_{\rm inner})$
value is $\sim3\times 10^{19}$ g s$^{-1}$
$-$
a factor $\sim30$
higher than observed.
HW employed an early, experimental version of
Cannizzo's time dependent disk code (from 1984),
which may have led to numerical problems.
For instance,
  they were forced to adopt an inner radius
$\sim10$ times too large.
In this paper we use a much more modern version
of the code and find ${\dot M}\sim 10^{18}$ g s$^{-1}$
as observed.

(iv) In running long term models
with the parameters given in CCL so as to cover
complete cycles of quiescence and outburst,
we find that outbursts tend to be triggered
frequently at small radii,
    confirming the analytical
   estimates of Lasota et al. (1996).
Small outbursts occur every $\sim5$ yr, whereas
major outbursts that could be observed with typical
   all-sky monitors
on X-ray satellites
    occur every $\sim50$ yr.
The rise times for the outbursts are $\sim10$ times
longer than those commonly observed.
When we include some evaporation of matter
from the inner edge so as to produce a truncated
$\Sigma(r)$ profile in quiescence,
we find that the smaller outbursts
are
eliminated and the rise times associated with the
major outbursts are shortened,
but not enough to bring them into accord with 
observations.
Therefore we cannot
confirm the conclusions of HLMN.
  The  parameters for their system of interest,
  the microquasar GRO J1655-40,
are sufficiently different from A0620-00
that a direct comparison of our respective findings
   may not be justified.
   In our modeling we find  that, 
      if a given model
for the quiescence is not able
   on its own 
  to prevent
inside-out outbursts
from occurring,
then the addition of evaporation
as an outside agent will not influence
the $\Sigma(r)$ profile 
to the extent required.
We note that
the form for $\alpha$
arrived at by CCL
is really only a constraint on
$\alpha$ in the hot state (i.e., on the
upper stable branch of the ``S'' curve).
Were one to artificially lower $\alpha$
in quiescence
by, for instance, reducing the proportionality
constant $\alpha_0$,
it may be possible to force the matter
to keep from evolving significantly at
small radii and thereby prevent inside-out
outbursts from triggering.
Another  possibility
might be for the cooling front
to ``stall'' at some radius
$\sim10^9-10^{10}$ cm
so as to maintain a permanent
hot, inner disk.
This  would  deplete
  the $\Sigma(r)$  profile
   for small radii
    and produce quiescent X-rays
   at a level commensurate
   with that seen in 
   A0620-00 using a standard
  efficiency $\epsilon\sim0.1$.

For our standard
model plus evaporation
at a level ${\dot M}_{\rm evap}\sim10^{14}$
g s$^{-1}$,
  we produce outbursts with
about the right spacing and amplitude
  as observed
($t_{\rm recur}\sim50-100$ yr,
 outburst energy $\delta E\simeq3\times10^{44}$
ergs, accreted mass $\delta M\simeq5\times10^{24}$ g,
and   peak absolute visual magnitude
      $M_V\simeq+1$),
although the rise times are too slow.
    As noted earlier,
       Lasota et al. (1996)
point out that the standard
model for the limit cycle instability
in the BHXBs produces outbursts which recur
too frequently
    because the  inner disk radius
     is quite small,
and they argue for evaporation
to truncate the inner disk.
If evaporation at a level $\sim10^{14}$ g s$^{-1}$
is correct,
then the efficiency of accretion in the low
state must be $\epsilon\sim10^{-4}$
(Narayan et al. 1996)
to account for the observation
of McClintock et al. (1995) of A0620-00
in quiescence.
Following this one step further,
because the efficiency in quiescence
of the neutron star SXTs
is still $\epsilon\sim0.1$,
there should exist a difference
of $\sim10^3$ between the quiescent fluxes
in BHXBs and neutron star binaries.
This does not appear to be the case
(Chen, Shrader, \& Livio 1997, see their Fig. 14).
Hence, while evaporation
appears to be attractive in some ways,
it may prove to be problematic in others.

(v) The local values of the
   irradiation temperature
   are significantly less than the local
values of the disk temperature
in our standard model.
King \& Ritter (1998)
argue quite the opposite,
     that irradiation
is in fact strong during
the outbursts of the BHXBs.
They propose that the slow,
exponential time scales for decay
seen in A0620-00 and other
   similar  systems
are caused by the inability  of the 
cooling front to form and shut off
the flow of matter
onto the BH.
We may estimate the strength
of the irradiation
during outburst in A0620-00
in several ways.
Warner (1987) found an empirical
relation between absolute visual magnitude
   $M_V$ and orbital period 
   $P_{\rm orbital}$
for dwarf novae at maximum light in
an outburst.
A similar relation was found
by van Paradijs \& McClintock (1994,
   hereafter vPMc)
for outbursts in BHXBs and neutron star
binaries.
Their relation was
between $M_V$ and 
  $(L_X/L_{\rm Edd})^{1/2} P_{\rm orbital}^{2/3}$.
The underlying cause for the Warner relation
is thought   to be the fact that
the  mass which
can be stored up in quiescence in the limit cycle model
is bounded from above 
     by a critical value
$\int 2\pi r dr \Sigma_{\rm max}$,
and this value scales with the size of the system
(C93b; Warner 1995; Cannizzo 1998a).
   There seems to be general
   agreement that the triggering mechanism
   for outbursts in both classes of systems
  is the  limit cycle instability,
  therefore the ``maximum mass'' concept
   should apply to both.
By comparing the $M_V$ values between the Warner and vPMc
relations, one sees that  the SXTs are intrinsically
brighter than the dwarf novae.
VPMc argue that the form of their relation
implies a strong X-ray heating effect which leads
to the bulk of the optical flux, through reprocessing,
in the SXTs.
In going between the dwarf novae
     and  the neutron star
SXTs, both of which presumably have $\sim1\msun$
accretors, the comparison at a given orbital
period should be direct
   and allow a pure estimate
    of the strength of the 
  irradiation, because the sizes
of the two systems
  and hence the surface areas of the
   accretion disks   should be similar.
For the BHXBs, however, one also needs to take 
into account the increased size of the disk,
at given orbital period, due to the presence
of a more massive central object.
If we compare the Warner and vPMc relations
at the approximate orbital period of A0620-00,
we see that $\Delta M_V\sim3$ mag.
For $P_{\rm orbital}$ constant,
the orbital separation scales
as $M_1^{1/3}$
(if $M_1>>M_2$),
so the radiating area of the disk
which gives rise to the optical
scales as $M_1^{2/3}$.
In going from a $\sim0.7-1\msun$ WD
to a $\sim7\msun$ BH,
we increase the disk area
and hence optical flux  by $\sim7^{2/3}-10^{2/3}$
$\sim4-5$,
which reduces the difference $\Delta M_V$
between the Warner and vPMc relations 
for an A0620-00 type system by $\sim2$ mag.
Therefore $\Delta M_V$(corrected)$\sim1 $ mag.
This comparison is admittedly crude
because there may also exist a correction
due to an effective temperature difference
(J. van Paradijs 1997, private communication).
We can obtain a more direct estimate simply by comparing
our computed $M_V$ for
A0620-00 with that observed
  in the 1975 outburst.
  The shortfall in optical flux
in the models which ignore irradiation
   must be due to the reprocessing of
  X-ray flux.
  Therefore by determining the deficit,
we learn how much of the optical flux comes
from reprocessing, and how much is due to
  viscous dissipation.
  This exercise has already been carried out
  by previous accretion disk limit cycle modelers
  specifically for A0620-00
  using time dependent models which
did not include the effects of irradiation.
  HW
  modeled the $B$ band flux and
determined $\Delta M_B\sim1$ mag
at maximum light (see their Fig. 3),
and MW
modeled the $V$ band flux
and found no difference between
their computed value and the observed
value at maximum light (see their Fig. 10).
In our model we find, for a face-on disk,
$M_V\sim+1$ at  maximum light,
whereas the observed value,
corrected for extinction,
is $+0.7$ (vPMc).
If we increase the inclination of our system to
$\sim70^{\circ}$ (Haswell et al. 1993),
our system becomes fainter and
we add   $\sim1$ mag to our $M_V$(peak) value,
thereby giving $\Delta M_V\sim1$ between theory and
observation, in line with the other estimates.
Finally, in comparing our nonirradiated
and irradiated models,
we find at the peak of the outburst that
$T_{\rm irradiation}/T_{\rm effective}\sim0.3-0.4$
for $n=2$ thought to be applicable
   to the BHXBs.
VPMc show  that, for the temperature
regime of relevance for the SXTs,
the optical flux scales roughly as $T_{\rm eff}^2$.
Therefore were we to self-consistently include
the effects of irradiation in computing $M_V$,
we would produce more optical flux by 
a factor $\sim1.3^2-1.4^2\sim2$,
which would subtract $\sim1$ mag from
the inclination corrected value $M_V\simeq2$,
bringing it back  into accord with the observed value.
In summary, it seems that the contribution of
the reprocessed X-ray flux to the optical flux
at peak light in A0620-00 is $\sim1$ mag,
meaning its relative contribution is $\sim50-70$\%.
This can be produced in the  time dependent disk
models by a slight increase of the effective
temperature
in the inner part of the hot, viscous accretion
disk.
Were the entire disk to be strongly irradiated
out to $\sim0.7$ of the Roche lobe of
the primary so as to prevent the cooling wave from forming,
           as advocated by King \& Ritter (1998),
the optical flux would be 
overproduced with respect to what is observed
in A0620-00.

We may quantify this last statement
using the subroutine
of our time dependent code
which calculates the $V$
band flux.
If the entire disk
is irradiated so that
the cooling front is unable to
form and propagate, that
means the effective
temperature must be at least
$\sim9000$ K.
The irradiation flux falls off
as $r^{-2}$,
so the radial distribution
of the effective temperature,
if it is dominated
by irradiation flux,
is $r^{-1/2}$.
Inputting the law $T_{\rm eff}(r) = 9000$ K
$(r/r_{\rm outer})^{-1/2}$,
where $r_{\rm outer}=1.5\times 10^{11}$ cm,
we find $M_V=+0.45$.
For a moderate
inclination, this becomes $\sim1.5$.
   If the entire
exponential decay
observed in $V$ during the 1975 outburst
of A0620-00 were irradiation dominated,
this is the faintest
the disk could become.
At the end of the exponential decay,
however, one saw $m_V\sim14-15$
(Lloyd et al. 1977),
 corresponding to an absolute
 $V$ magnitude  $M_V\sim3-4$ (vPMc)
$-$ 
fainter than seemingly allowed
by the irradiation model.
 As a  test of  the zero point of the $V$
calibration, we input $T_{\rm eff}
=T_{ \odot} = 5770$ K
(constant with radius) and
 $r_{\rm outer}=R_{\odot}=7\times10^{10}$ cm,
and find $M_V=+4.86$.
The accepted solar value is $M_V=+4.83$
(Allen 1976).
 King \& Ritter (1998) 
  point out that the constraints
  on the functionality of $\alpha$
derived from arguments like those presented
in Cannizzo et al. (1990)
  do not apply for disks in which
   the midplane  temperature is determined
  not by viscous heating but rather
   by  irradiation.
For the midplane temperature
  to be strongly influenced,
a rather large irradiation is needed
(Tuchman et al. 1990).
Our estimate that $T_{\rm irr}\sim10^4$ K
at the outer disk edge may be unrealistically low.
During the time in which the optical flux
decreased by $\sim2-3$ mag
in the 1975 outburst of A0620-00,
the X-ray flux fell by a factor $\sim10^3$.
It is not clear how the strongly irradiated
midplane would be able to maintain
a constant temperature $T_{\rm mid}$
during this period of declining
irradiation so as to lead to the kinematic viscosity
coefficient $\nu$ being constant
with time in the outer disk.
The constancy in $\nu$ is needed
in order to produce an exponential decay
of the visual and X-ray fluxes
in the model of King \& Ritter 
(Mineshige et al. 1993).

  In this work we have chosen to
  concentrate on the most basic
 features of the BHXB outbursts: 
  the recurrence time scale,
   the $\sim30$ d exponential
  decay, and the $\la3$ d fast rise.
 We do not directly address the issue
  of the secondary maxima or reflares
   which are frequently  seen.
  Based on some of our findings,
however, we may discuss previous work on the origin
 of the reflares.
  These are resurgences in the X-ray flux
  which are also seen to some extent in the
optical,
  during which time the flux increases 
  by a factor $\sim2$ before resuming its
decline.
  The decay properties after the reflare
  are much as before, so it is as if
  the light curve were displaced upward
  by a factor $\sim2$
           beginning at a point 
 $\sim60-70$ d after the outburst starts.
   Several works
    have appeared
 concerning the reflares
  (Chen, Livio, \& Gehrels 1993,
    hereafter CLG;
   Augusteijn, Kuulkers, \&  Shaham 1993;
  Mineshige 1994; K96;
  Kato, Mineshige, \& Hirata 1995,
   hereafter KMH; 
   IMK;
   Kuulkers, Howell, \& van Paradijs 1996, 
   hereafter  KHvP;
   for a review see 
     Wheeler 1998).
  One feature which some of the theories of the reflare
 share is that they involve irradiation,
either of the secondary star so as to induce additional
  mass overflow which powers an increase in the
luminosity by virtue of the extra material added
to the disk (CLG; Augusteijn et al. 1993),
  or irradiation of the disk so as to reflect
the cooling wave at  one or two grid points 
  as a heating front and momentarily rebrighten
  part of the disk (K96).

In the context of the theories
which rely on irradiation,
  it is interesting that
   the reflares seem to be uniquely
associated with the BHXBs (Wheeler 1998).
  Reflares are not seen   in the neutron star
SXTs such as Cen X-4 and Aql X-1.
  According to our estimates of the
  strength of irradiation,
  and that of previous workers
  (HW; MW;
  van Paradijs 1996), however,
   irradiation is weak in the  BHXBs.
  If reflares are unique to BHXBs,
  and if irradiation is weaker
   in BHXBs than in neutron star SXTs,
it seems rather unlikely that irradiation
  can be the BHXB specific parameter
  which causes reflares  to occur
   only  in the BHXBs.
  This makes it difficult
  to accept any of the irradiation-based
      models.
       Another consideration is that,
         whatever the physical mechanism is,
  it must be such that it leads
  to a delay  time between onset
  of the main outburst and secondary reflare
  that scales with orbital period.
   The fact that the reflare in GRO J0422+32
occurred $\sim45$ d after 
primary maximum, whereas
in A0620-00 it occurred $\sim55$ d  after
primary maximum, led CLG to
predict an orbital period less than 7 hr
for this system.
The period was later found to be 
  5.1 hr (Chevalier \& Ilovaisky 1994,
Orosz \& Bailyn 1995).

What unique property of BHXBs $-$
something which would not be true for the
neutron star SXTs $-$
 exists that could account for the
  reflares?
This has already been
answered by many authors
$-$
it is the extreme value
of the mass ratio $q=M_2/M_1$
(Bailyn 1992; KMH; IMK; KHvP).
This ratio is much smaller for BHXBs than for
neutron star  SXTs,
and this can lead to resonances occurring within
$\sim0.7$ of the Roche lobe of the primary
wherein the disk is constrained
by tidal truncation to exist.
One observes superhumps (SHs) during
          superoutbursts (SOs) in the
 SU UMa class of dwarf novae (Warner 1995).
The SHs are thought to be caused by the
fact that the 3:1 commensurability resonance
with the binary
      orbital period lies within the accretion
disk for $q\la0.25$ (Whitehurst 1988, 1994, 1995),
but lies beyond the point of tidal truncation
 for $q\ga0.25$.
In fact, SHs are only seen in short orbital period
dwarf novae
for which $q\la0.25$.
Whitehurst showed that when the condition $q\la0.25$ is fulfilled,
the outer disk acquires an elliptical shape
which precesses in the corotating frame
of the binary,
and this precession causes the distance
through which the mass stream must fall
before striking the outer disk
to vary, thereby modulating the light
 curve on a time scale equal to the
orbital period plus  a few percent.
Osaki (1989) combined the disk instability model
with a proposed ``tidal instability'' which he
envisioned to occur whenever Whitehurst's
condition for superhumps becomes satisfied.
Osaki's thermal-tidal (TT) instability allows
for superoutbursts in the SU UMa stars
by imagining that,
whenever the outer edge of the disk
expands beyond the 3:1 resonance in the disk,
the increased tidal torque on the disk
enforces a contraction of $r_{\rm outer}$
by $\sim10-20$ \%.
The mass that was present in the disk at the start
of the outburst is confined to a smaller volume,
and the increase in $\Sigma(r_{\rm outer})$
relative to $\Sigma_{\rm min}(r_{\rm outer})$
produces an ``over-filled'' disk that
prevents the cooling wave from starting
until a sizable fraction of the disk matter
has accreted onto the central object.
The motivation for the contraction
of the outer disk was not obvious,
based on Whitehurst's time dependent
calculations,
and led to criticisms (Whitehurst 1994; Cannizzo 1996a).
Recent work by Murray (1997)
     which utilizes a more advanced
SPH model than was available to  Whitehurst,
however, does in fact show a contraction
of the outer radius which is 
coincident with the growth into the nonlinear regime
of the elliptical disk induced
by the 3:1 resonance. This gives credence to Osaki's model.

What does this have to do with the reflares in
the BHXB outburst light curves?
A potentially key observation is that,
in the BHXB GRO J0422+32,
SHs appeared in the outburst at about the same
time as the  reflare,
   thereby arguing for a direct connection
between the nonlinear outcome of the
3:1   instability $-$ namely the formation of an eccentric
  disk and concomitant SHs $-$
and the reflare (KMH, KHvP).
The crucial difference which
     the global effect
the disk contraction has in going from
accretion disks in dwarf novae
to those in the BHXBs has to do with
the growth time to reach the  nonlinear regime
of the tidal instability.
Several workers 
  have suggested a direct connection
  between the SHs and the reflares
  (KMH, IMK, KHvP),
  namely,
  that it is the contraction
  of the disk in Osaki's TT model
 that leads to the reflares.
IMK argue that 
the 3:1 resonance growth time
varies as $t_{\rm grow}\sim1/(\Omega[r_{\rm outer}])$,
  where $\Omega(r_{\rm outer})$ is the Keplerian
  frequency at the outer disk edge.
They point out that $t_{\rm grow}$
is about 30 times larger for the BHXBs than 
for the SU UMa stars.
For normal outbursts which turn  into SOs
in the SU UMa stars by virtue of the
expansion of $r_{\rm outer}$ beyond
  the 3:1 resonance radius,
the time $t_{\rm grow}$(SU UMa)$\sim1-3$ d,
so that one sees in the light curves 
first
 a rise to maximum associated
 with the normal outburst, then a brief dip,
and then a rebrightening to maximum as the forced
contraction of the accretion disk becomes active
  and the disk gas gets confined to a smaller radius.
This constitutes the formal start of the SO,
 marked subsequently by a long term
slower-than-exponential decline that
typifies decays  from maximum light in which
the cooling front is unable to propagate.
If 
  $t_{\rm grow}\sim100$ d
for the BHXBs,
  this would have important
consequences for the interaction of the
contraction phase
with the inward propagation of
the cooling front.
This time scale is about three $e-$foldings
for the time to deplete the mass of the hot part
of the  accretion disk,
or one $e-$folding time
 $|d\ln r_f/dt|^{-1}$
 for the cooling front
to move inward
 (CCL),
so the cooling front has already had ample time to
   travel inward a considerable distance from $r_{\rm outer}$
as defined by the tidal truncation,
and subsequently redefined by the increased
degree of tidal truncation due to the 3:1 resonance-induced
elliptical disk.
There is no opportunity for the cooling front
to be prevented from forming as in the SU UMa
SOs, and yet the gas in the accretion disk
 must be confined in some manner to a smaller value.
It is difficult to estimate the degree of this contraction
without doing a detailed calculation,
considering especially the fact that the contraction
occurs primarily in the outer,
cold portion of the disk.
One suspects that the brief period
 of augmentation in the local $\Sigma(r)$
 values would produce a reflection of the cooling wave
as a heating wave, which would propagate out some distance
and then reflect inward again as a cooling front.
  This is found in time dependent calculations (IMK).
The amplitude of the reflares observed
in the BHXB outburst light curves are
about a factor of 2 with respect to the X-ray flux
 observed just prior to the start of the reflare.
This would not require a large reflection
in order  to be  produced.
The reflares caused by the reflection
of cooling waves into heating waves seen in some
of the light curves in CCL showed
increases of $\sim10$ in flux.
(Their cause is thought to
be numerical, however, and not physical.)
In summary,
  the fast growth time for the 
3:1 instability in the SU UMa stars
  leads to an immediate contraction
in $r_{\rm outer}$ and consequent
over-filled disk which can only decay viscously
(since the cooling front cannot propagate),
  whereas the slow growth time
for the 3:1 instability in the BHXBs means
that the contraction, when it does occur,
  is too late to produce an over-filled
disk. Instead, one has only a minor
re-adjustment in $\Sigma(r)$ which produces
  a reflected transition front
  and resultant reflare.

IMK attempted to calculate
  models using
the TT instability
 for BHXBs,
  but various deficiencies
  in their work led to
  outbursts which did not resemble
  those observed.
  They use a low resolution logarithmic
  spacing $N\sim20-40$
 to cover a dynamic range 
   in disk radii  $r_{\rm outer}/r_{\rm inner}>10^4$,
they use a parametrization of $\alpha\propto r^{0.3}$
(C94) which was criticized
 as not being valid for BHXB parameters (CCL),
  and their model contains detailed ad hoc
 prescriptions for the functional form
  of the increase in the tidal torque
  associated with the TT instability
(both $\alpha$ and the torque proportionality
constant each have $\sim6$ adjustable parameters).
    The light curve
  they   present covering an outburst
  with the TT induced reflare
   shows a factor $\sim10^3$ increase
in $L_X$ in the reflare,
  rather than the observed factor $\sim2$ increase.
The decay characteristics are also
not as observed.
Nevertheless their model
  represents a first step
  toward  a complete theory of outburst
  plus reflare in a time dependent model.
Work is currently underway
 to activate and test the portion
of our code which allows for a variable
 outer disk radius so as to test the
global manifestation of Osaki's 
TT instability in the BHXBs.

\vfil\eject

\section {CONCLUSION  }

We have investigated
several aspects of the 
accretion disk limit cycle
mechanism as it applies to BHXBs.
Our main impetus is to extend
work reported in CCL by running
more extensive time dependent models,
cover complete cycles of outburst and quiescence,
and look in detail at 
several aspects of the model not
studied by CCL.
We find that the explanation
for the cooling front speed put forth
by VW seems to account for
the deviation  from steady state flow which we
observe.
In running complete cycles,
     we find
a strong tendency using the CCL $\alpha$
form for
    frequent  outbursts to
  occur which  begin at small
radii in the disk and    have slow rise times,
in contrast to the observations.
Evaporation of material from the inner
disk does not shift the surface density
distribution enough to decrease
   the rise times sufficiently,
  although it does increase
   the recurrence times for outbursts
  to an acceptable value
   (Lasota et al. 1996).
A separate issue is the necessity
of having some mechanism
distinct from the standard
limit cycle operating
in order to supply
material to accrete
onto the central object
in quiescence so 
as to provide
substantial X-ray and EUV radiation.
Finally, for our standard model which is meant to
represent a system like A0620-00,
irradiation of the disk does
not appear to be strong enough to
change the structure significantly.
This may pose a challenge
for workers that seek to invoke
irradiation to prevent the cooling
front from propagating.
Also, the apparent
weakness of irradiation
in the BHXBs relative to
the neutron star binaries
may present problems
for theories of the reflare
which are based on irradiation.

This paper is not intended
to be the final word in this subject,
but should only serve to pose additional
questions which must be looked into.
There is clearly much more work to be done.
In spite of our efforts we have failed
to produce a successful ``standard model''
for A0620-00 which reproduces not
only the  $\sim30$ d $e-$folding time exponential decay
and $\sim60$ yr recurrence time scale for outbursts
but also the fast rise $-$ $\sim3-5$ d.
We have purposely avoided introducing modifications
to the $\alpha$ form utilized by CCL
in order to provide continuity with that work.
It now appears, however, that it may prove
necessary to reduce $\alpha$ in the low state
below that which would naturally occur using
the CCL scaling. This could be carried
out for instance by reducing $\alpha_0$.
There may be other ways to
get the fast rise,
such as having a stalled cooling
front which would deplete $\Sigma(r)$
 for small $r$.

In this work we have only activated
the portion of the our code which calculates
the irradiation temperature
based on the central flux from the disk
and the local shape,
as  explained in Tuchman et al. (1990).
The next step is to activate
and test
the portion which also ``corrects''
the local disk temperature
   (both effective and midplane)
     based on the
irradiation temperature,
which would in turn affect the global
evolution of the disk (as in K96).
Since the irradiation
temperature does not even
approach the effective temperature
for parameters relevant
to A0620-00,
the kinematic  effect on the disk
   would be insignificant.
For the neutron  star systems,
  however,
  irradiation  will most likely 
   play an important role.

We have also chosen not to activate
the part of the code which treats
the variation of outer disk radius
(Smak 1984).
This may affect the nature of the
light curves
accompanying the outbursts,
in particular the rise times,
by virtue
of the mass swept up
at large radii
by the arrival of
low angular momentum
material
from the secondary star.


We acknowledge useful discussions
  with W.  Chen, 
  J.-M. Hameury,
   A. King,
   E. Kuulkers,
  J.-P. Lasota,
  C. Mauche, 
  J. McClintock
  R. Narayan,
   J. van Paradijs, 
   E. Vishniac, 
   B. Warner,
  and C. Wheeler.
   JKC
 was supported 
through the long-term 
scientist program under
 the Universities Space Research Association
(USRA contract NAS5-32484)
in the Laboratory for High Energy Astrophysics
at Goddard Space Flight Center.

\vfil\eject
\centerline{ FIGURE CAPTIONS }

Figure 1. The decay from maximum light for a disk initially
   entirely in the high state, but with 
$\Sigma(r)_{\rm outer}  < \Sigma_{\rm min}$ 
so as to ensure the cooling front starting.
We show the light curves in 
absolute visual magnitude $M_v$ ({\it top panel})
and the locally defined decay time constant
in days/magnitude ({\it bottom panel}).
The six  curves are
for $N=$ 21, 41, 81, 161, 321, and 501.
The last curve was started from an
initial condition with the disk
in the low state, but with 
a slight excess in $\Sigma$  at $10^9$ (cm)
to trigger the instability.

Figure 2. The variation in local flow rate and 
local disk thickness during a decay.
The plot shows
       $v_r/\alpha c_s$ versus $r$ 
({\it top panel})
and      $h/r$ versus $r$
({\it bottom panel}).
We show   eight time steps during 
the decay spaced 10 days apart
as the cooling front moves inward.
The x's indicate individual grid points.
The dashed line in the upper panel
shows the locus of grid points
$i_{\rm hot}$
which lie at the outermost
point of the inner, hot disk.
The dashed line in the second panel
shows the values of $h/r$
associated with $\Sigma_{\rm min}$,
and the dotted line gives
conditions at $\Sigma_{\rm max}$.

Figure 3. The radii at which the local
flow speed 
  (i) goes to zero
and 
  (ii) equals
the standard value
expected in a steady disk $\alpha c_s(h/r)$.
Both are given
      in units of the radius
of the hot portion of the disk
$r(i_{\rm hot})$.
The eight time steps shown correspond
to those depicted in Fig. 2.

Figure 4. The local time constant
  associated with the decay of the
optical flux for eleven values of
$n$, where $\alpha=\alpha_0(h/r)^n$.
The value of $n$ is 1.50, 1.52,..., and 1.70.
To keep the value of $M_V$ within a common
range, the proportionality constant
$\log\alpha_0$
is varied in steps of 0.04 from 1.7 to 2.1.

Figure 5. The variation of properties
associated with the transition front
during an outburst which is triggered
at a small radius.
We show the number of grid points
with the front $N_f$,
the fractional width of the front
$w/r$, and the ratios
$\sqrt{h/r}$ and $h/r$
evaluated in the outermost
radial zone
within the hot state.
Within the
rise of the outburst
$w/r\sim (h/r)^{0.75}$,
whereas during the decay of the 
$w/r\sim (h/r)^{0. 5}$.

Figure 6. The dependence of the
   outburst profile on the radial
location of triggering.
The panels show
the initial low state configuration
({\it top panel}),
and the resulting light curves
({\it bottom panel}).
The three small Gaussians in the top
panel are representative of the
narrow surface density enhancements
which are added to the broad,
underlying configuration,
in order to trigger the outburst.
The functional forms of the narrow
and broad Gaussians are given in the text.

Figure 7. The variation of the surface
  density with radius and time
for inside-out outbursts and outside-in
outbursts.
We show a model run for $\log r_{\rm center}$(cm)$=9$ 
and 
            for $\log r_{\rm center}$(cm)$=10$.
The dashed curve gives the initial state,
which is entirely along the lower branch
of the $S$ curve,
and the dotted lines correspond to
the local maxima and minima in $\Sigma$
from the steady state scalings.
We show  30 d of evolution,
with a 5 d spacing between each curve.
In the upper panel there
are six sharp spikes
evident between
$\log r$(cm)$=10$ and 11
showing the outward movement of
the heating wave at
$t$(d)$=$ 5, 10, 15, 20, 25, and 30.
In the lower panel there are also
six spikes in a similar radial range
showing the outward movement,
as well as five spikes
between $\log r$(cm)$=9$ and 10
showing the movement of the inner edge
of the heating front at $t$(d)$=$ 5, 10, 15, 20, and 25.
For $t$(d)$=$ 30,  the hot region
extends
all the way down to $\log r$(cm)$=7$.

Figure 8. A series of outbursts
spanning several cycles of quiescence and outburst.
The $y-$axis shows the 
log of the mass accretion
rate from the innermost grid point
in g s$^{-1}$,
and the $x-$axis
gives the time in yrs.
The top panel represents
the standard model
without evaporation.
For the
    next three panels we
remove material from the
inner edge of the disk
at the rate given in
the text,
with the
normalization constant
given in the text
multiplied by
$10^{-x}$,
where $x=4$, 3, and 2
({\it top to bottom}).

Figure 9. The time history of the
  mass of the accretion
disk accompanying the four light curves
shown in Fig. 8.

Figure 10.  One major outburst
   taken from each of
   the four panels
  in Fig. 8.
  The values of $x$ 
   are given beside 
   each curve.
Smaller $x$ values correspond to
larger evaporation.

Figure 11. The evolution of $\Sigma$
and $T$
  accompanying an outburst
in the third panel of 
   Fig. 8.
Ten time steps are shown,
with the first corresponding
to
just before an outburst begins.
There is a spacing of 0.1 yr
between curves.
The high temperature
at the inner disk edge
is an artifact of our
imposition that $\Sigma(r)\ge1$
g cm$^{-2}$, which prevents the inner
edge from cooling  completely.

Figure 12. The evolution
of $\Sigma$ and $T$
continuing from Fig. 11,
showing the quiescent 
period following the outburst.
The first time step
corresponds to
the last step in Fig. 11.
Four time steps are shown,
and the spacing is
 10 yr.

Figure 13. The ratio of 
the local irradiation temperature
to the local effective temperature
for 600 days of evolution,
shown at 6 d intervals,
for $n=1$.
The drop-off at
large radii moves inward 
as time progresses.

Figure 14. The ratio of 
the local irradiation temperature
to the local effective temperature
for 600 days of evolution,
shown at 6 d intervals,
for $n=2$.
The drop-off at
large radii moves inward 
as time progresses.

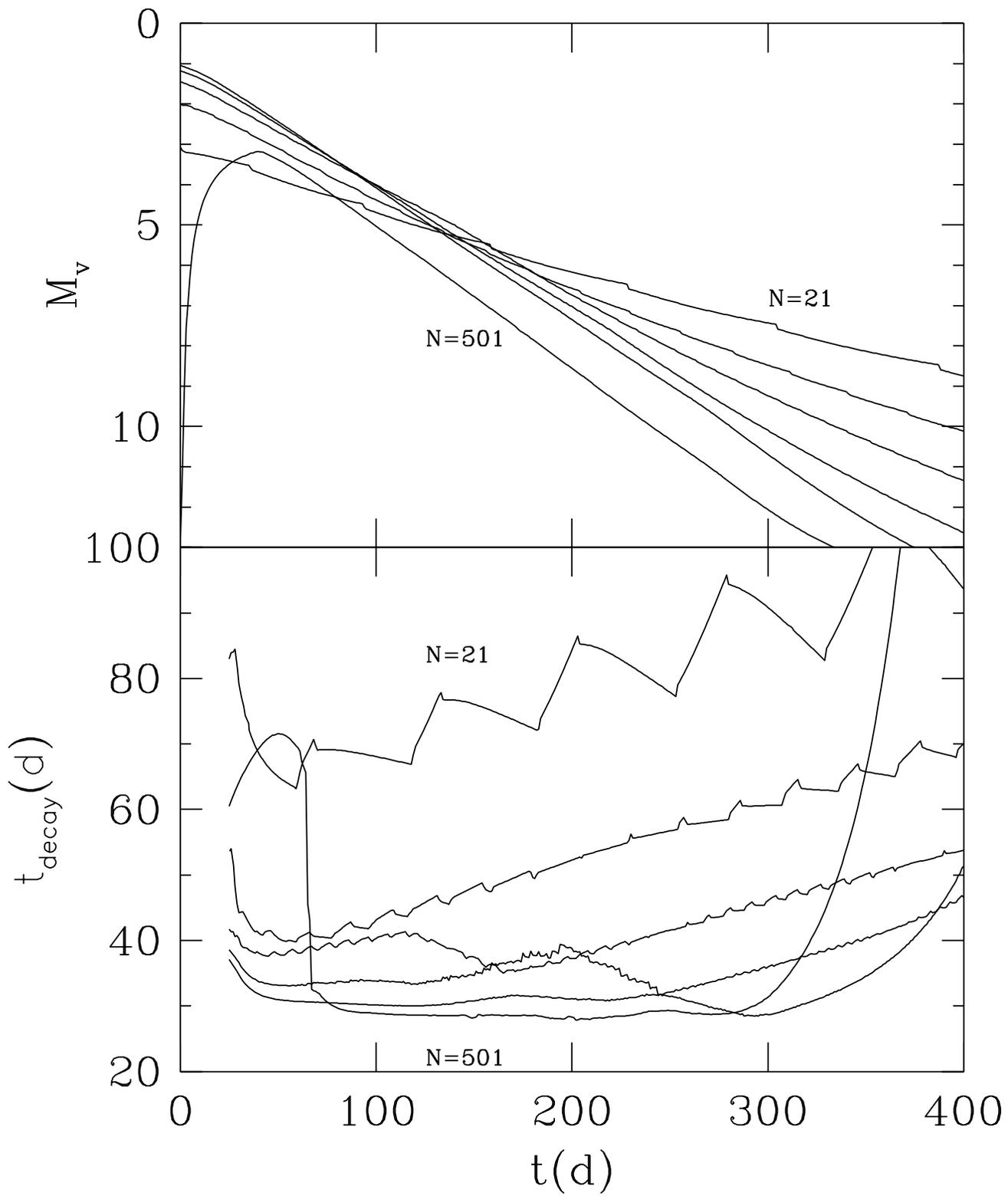
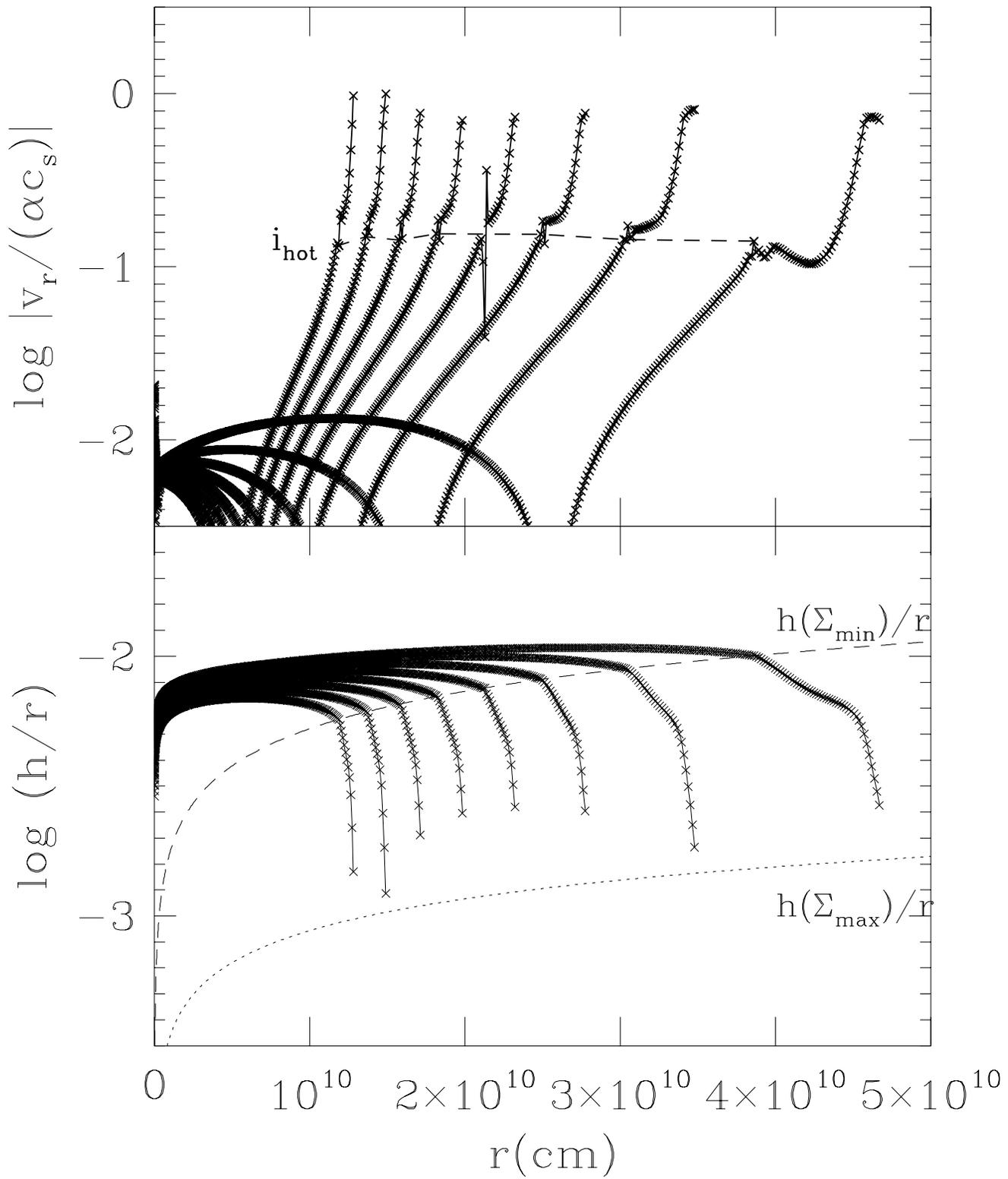
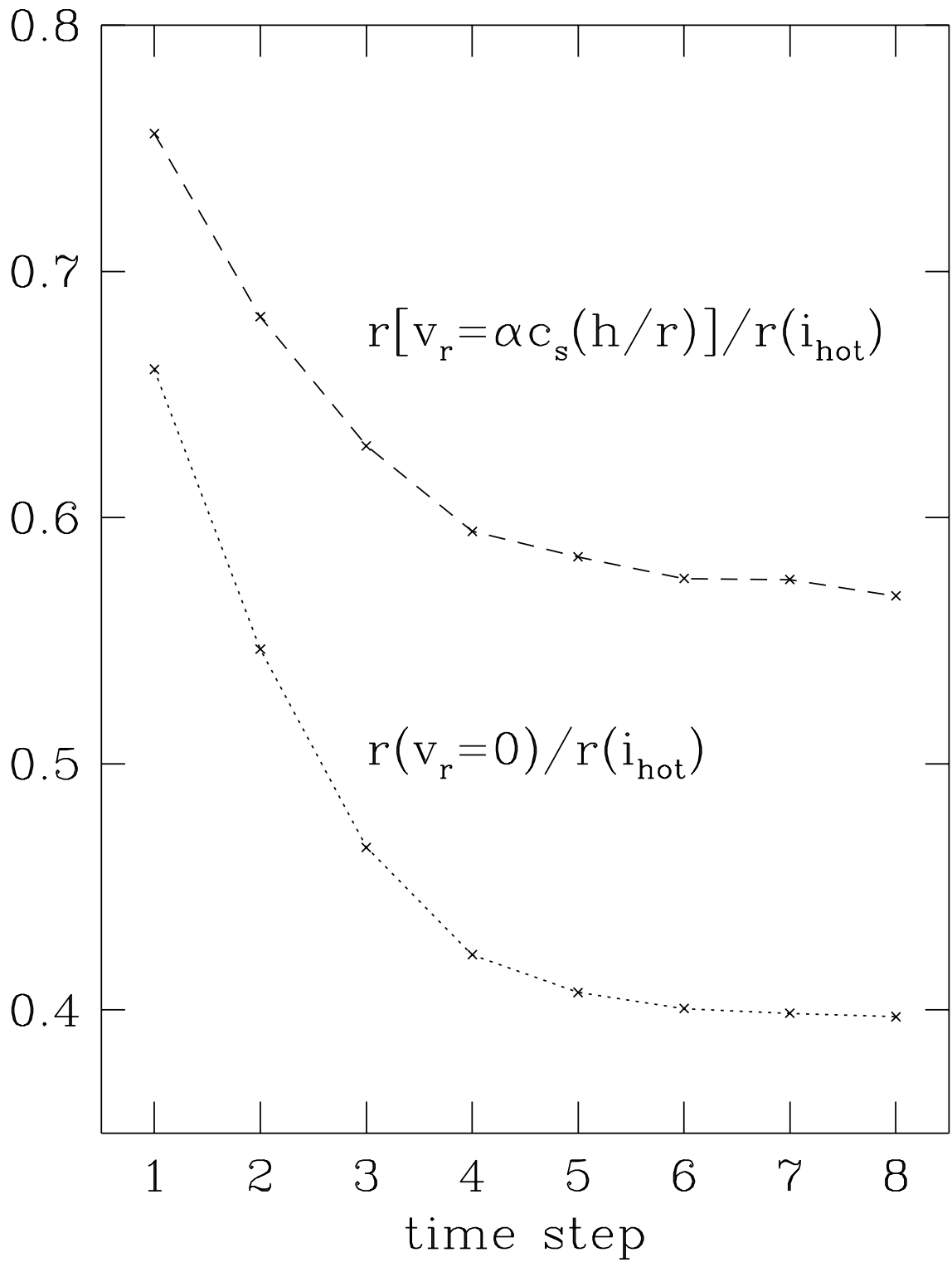
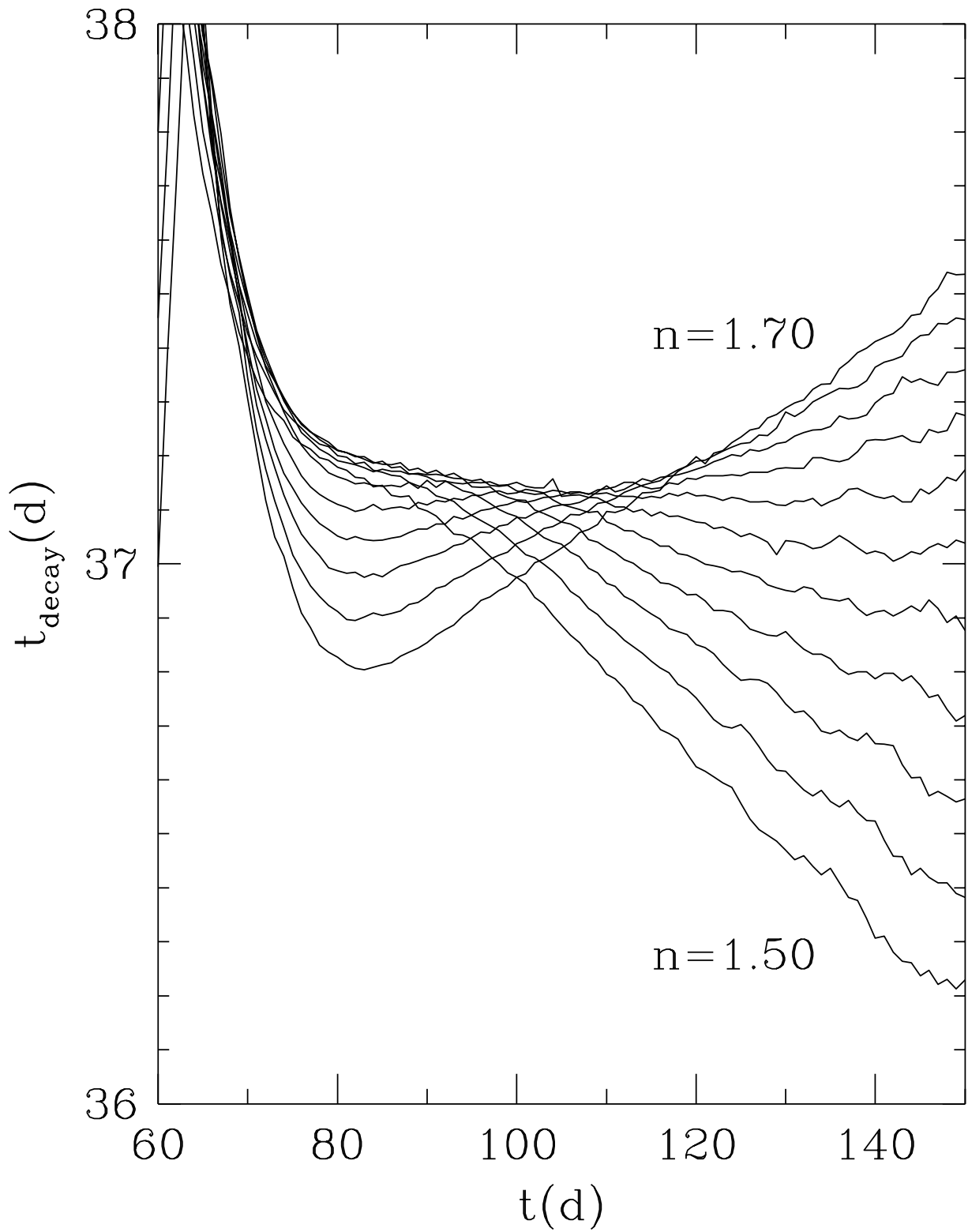
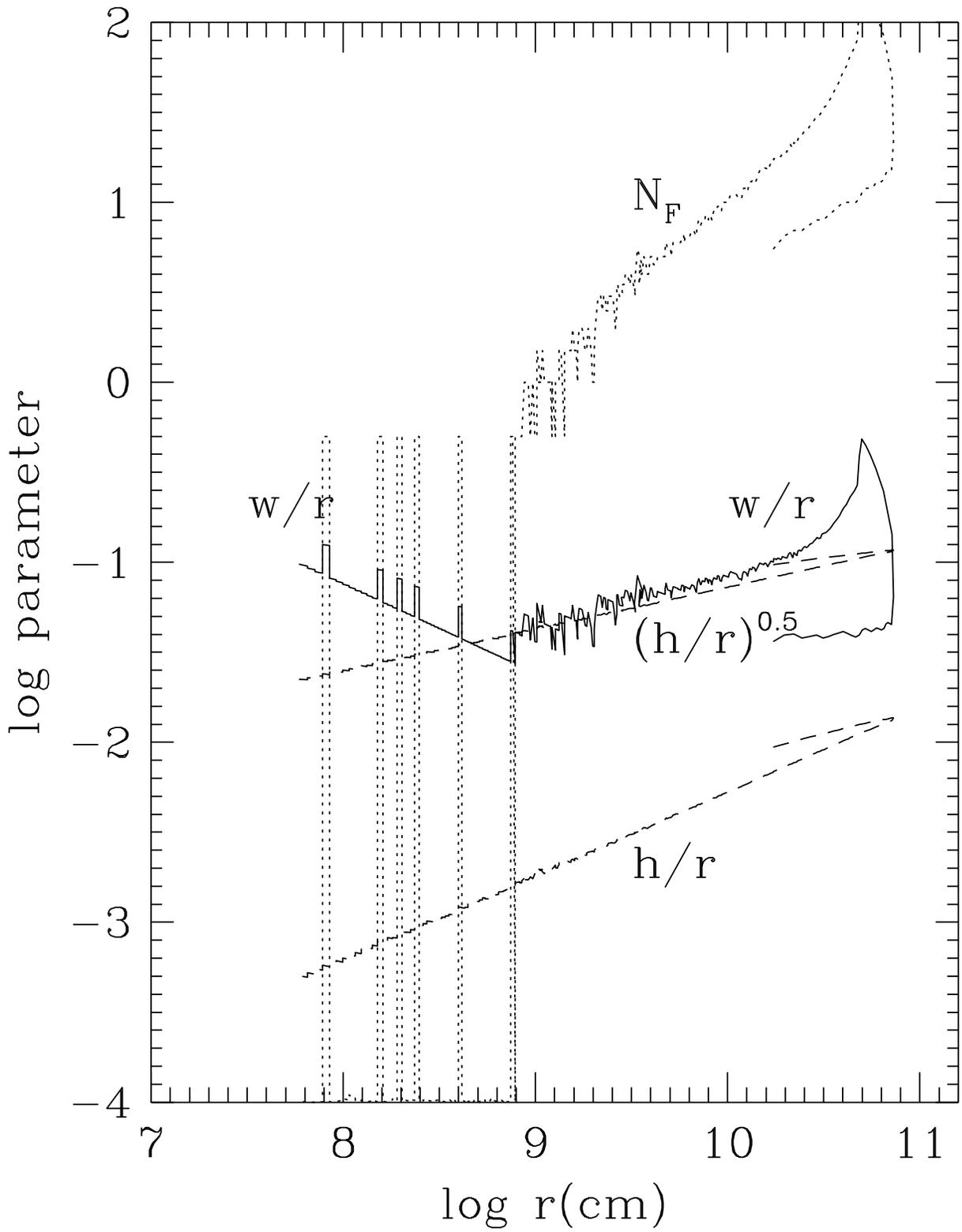
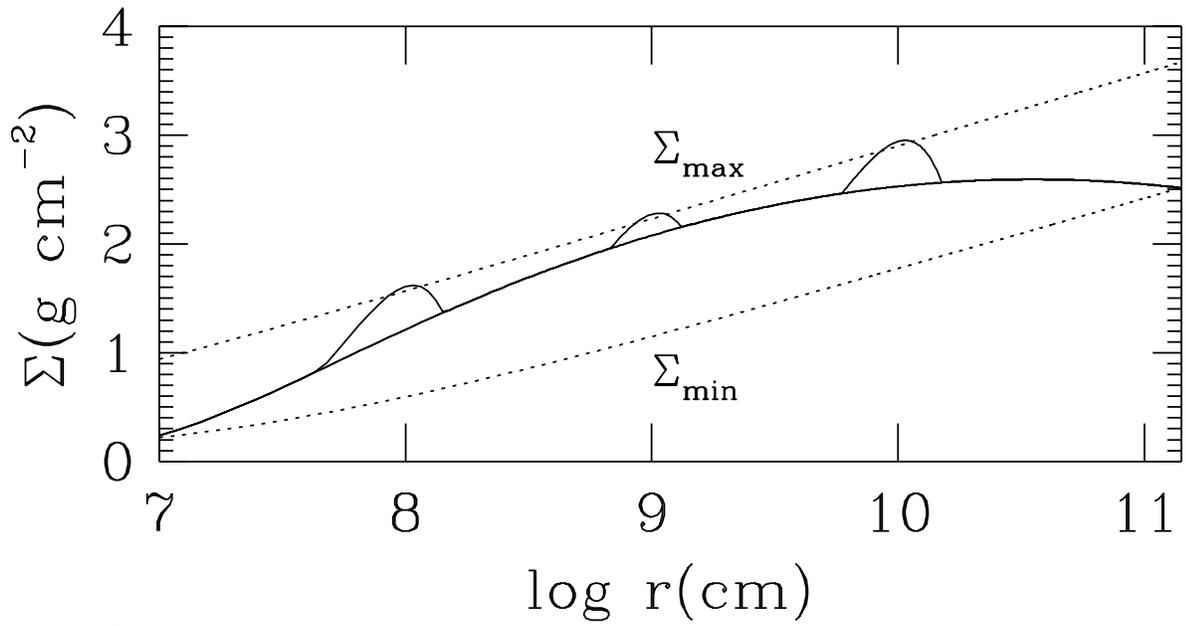
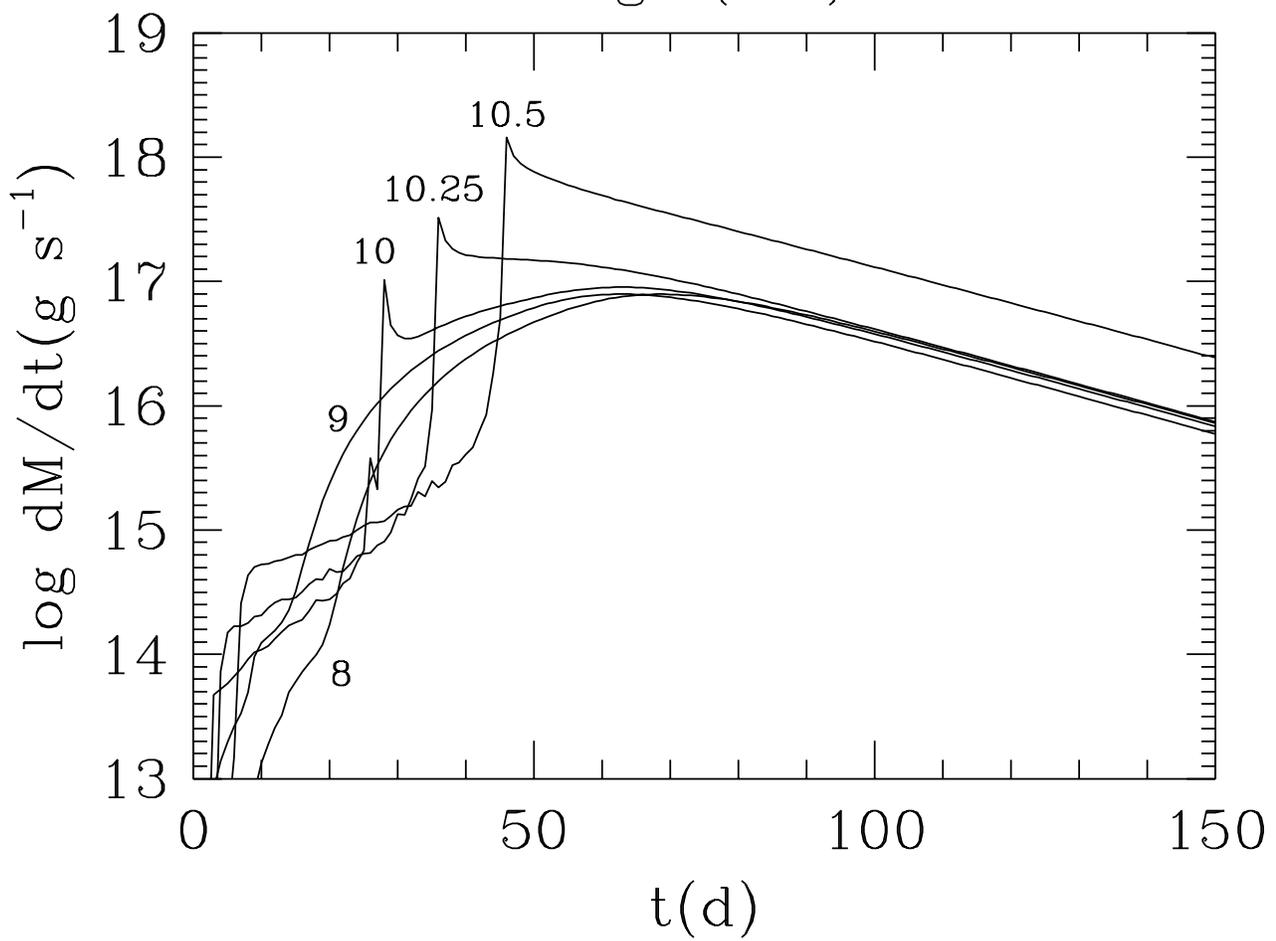
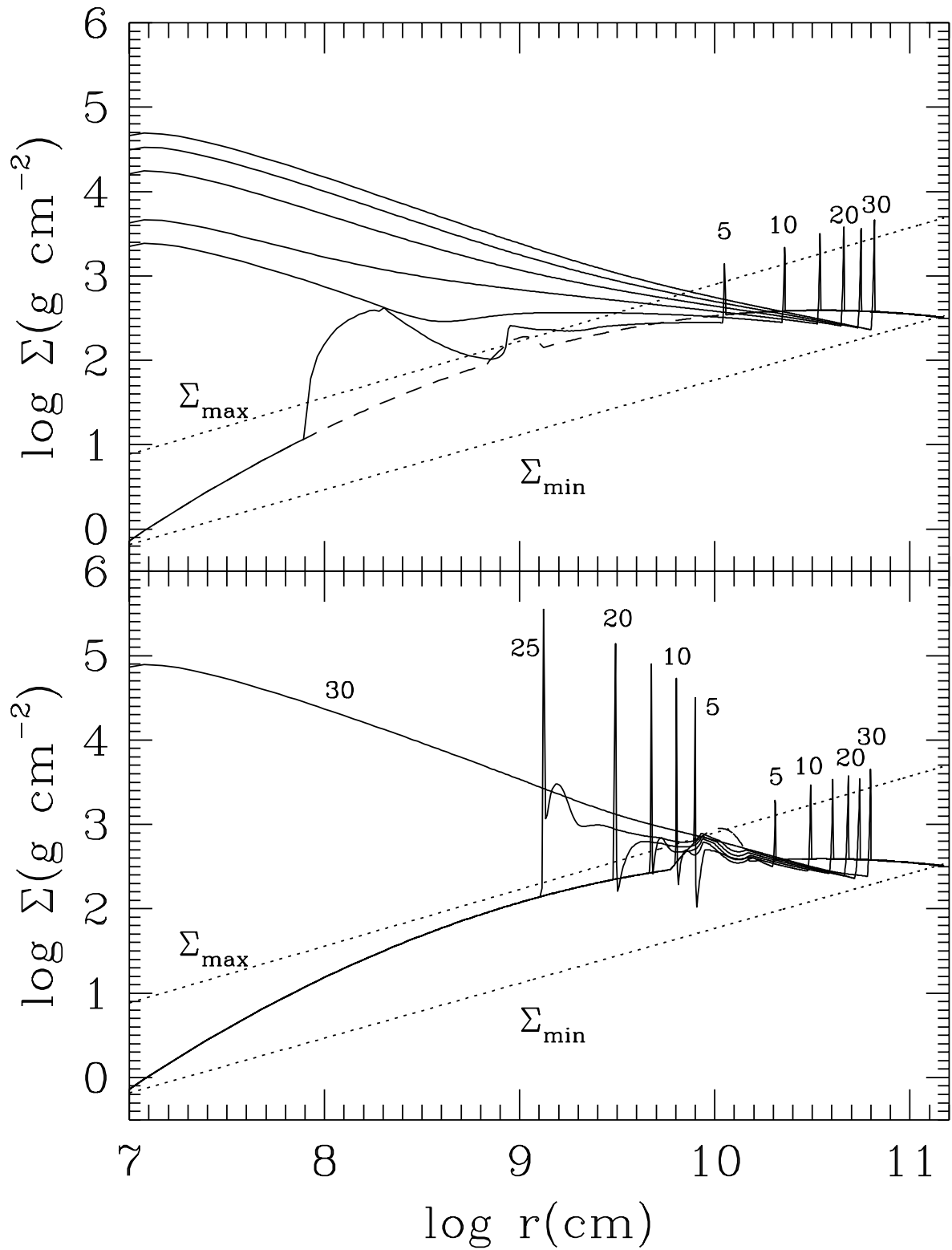
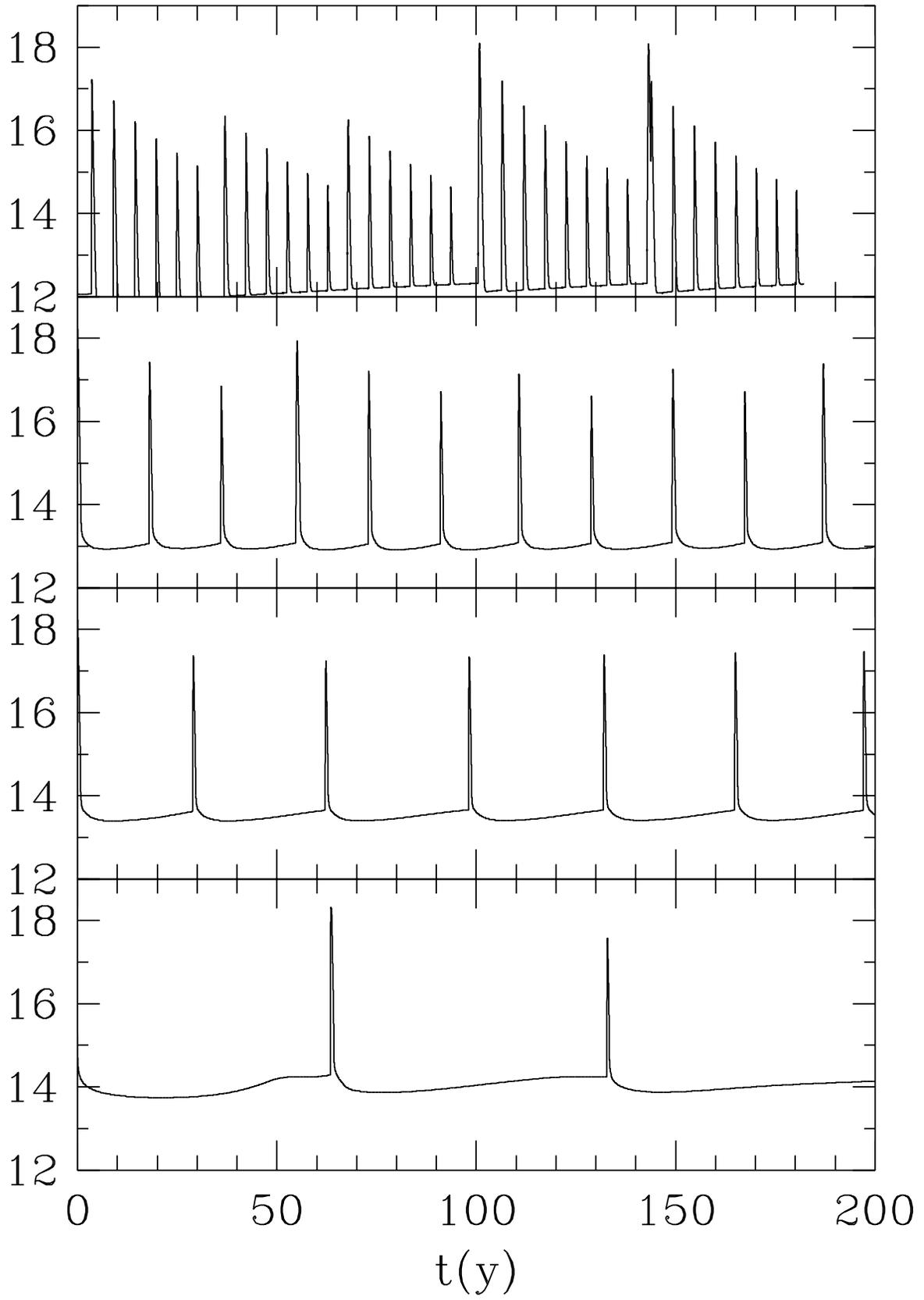
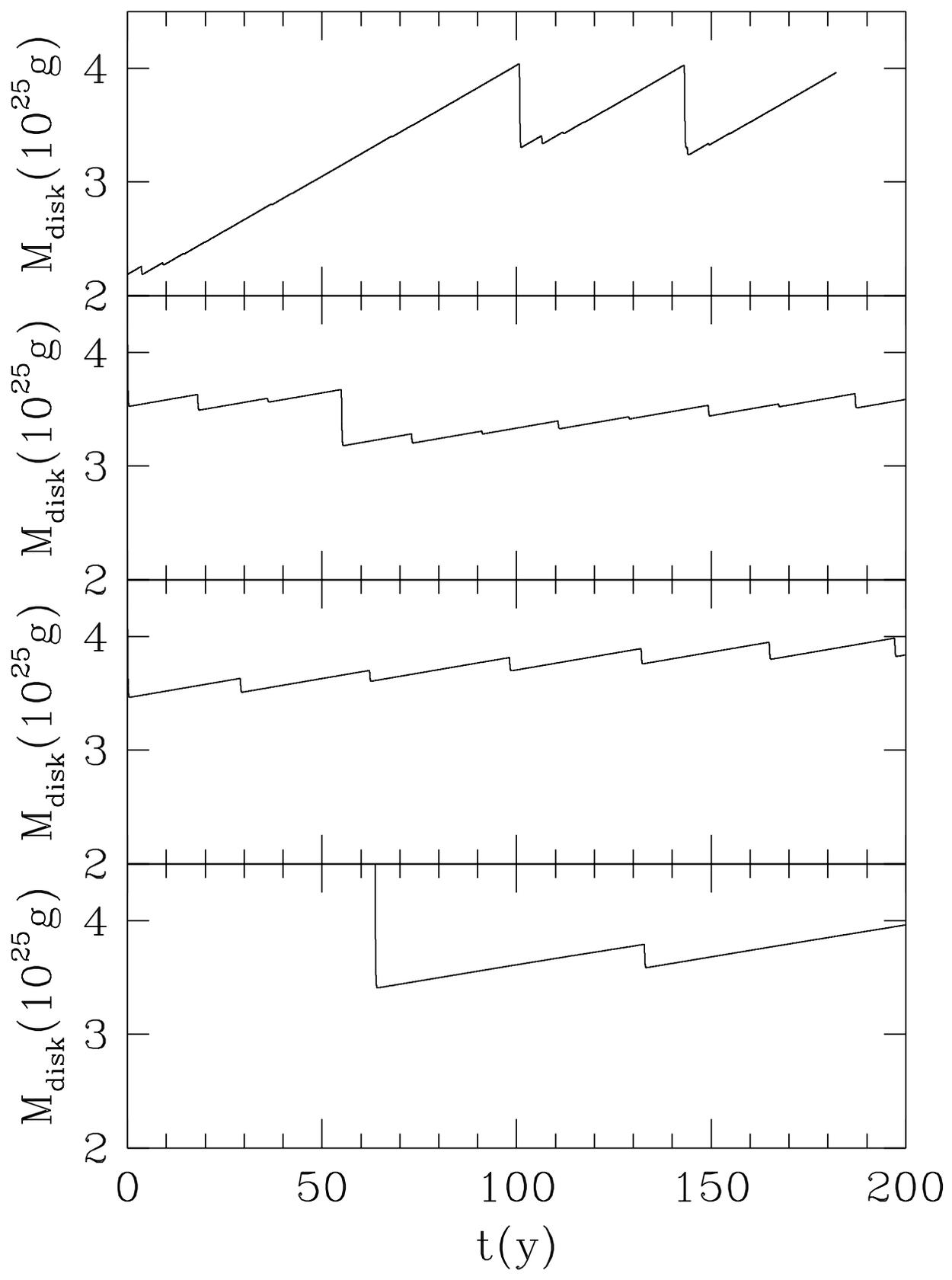
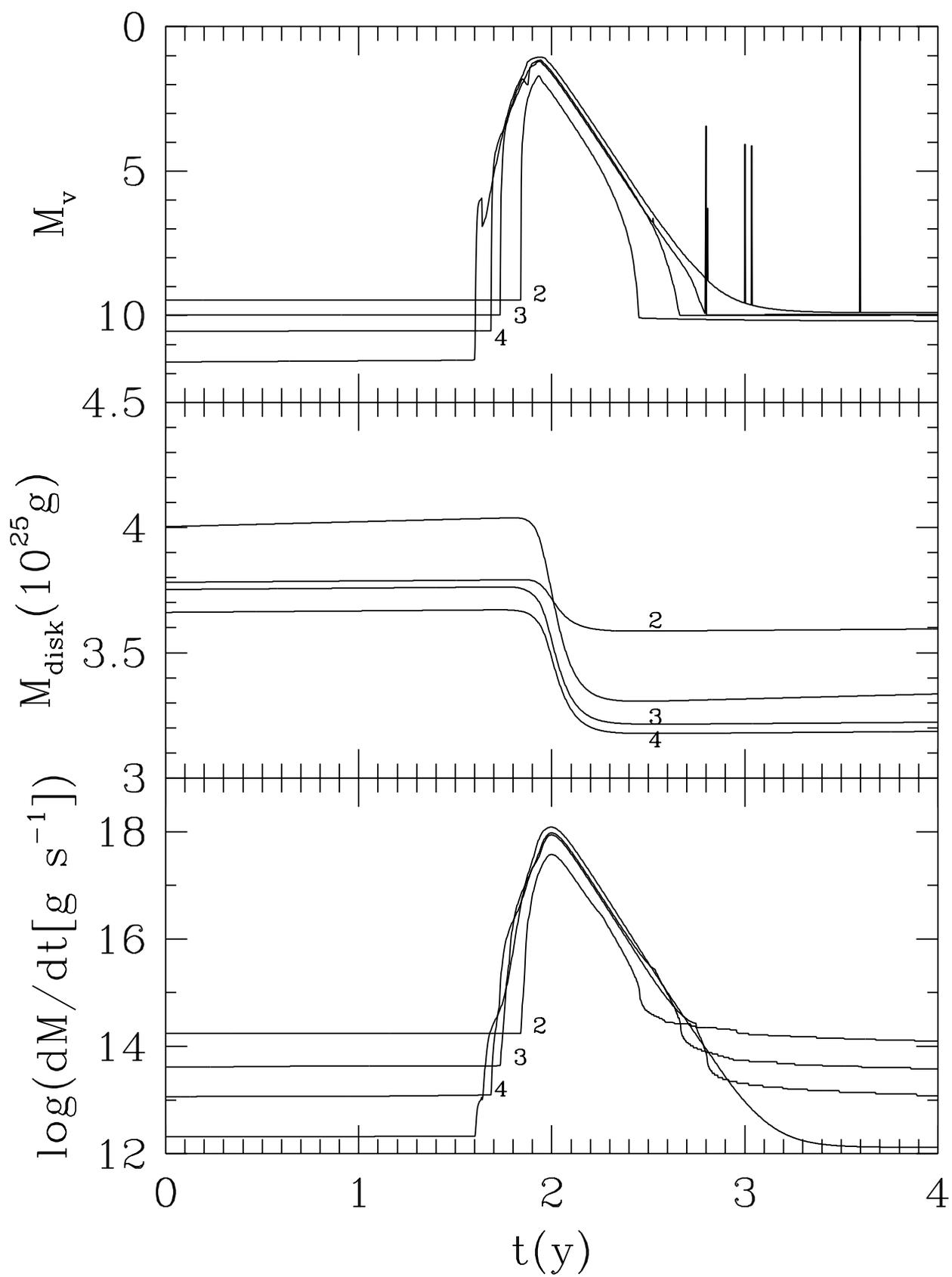
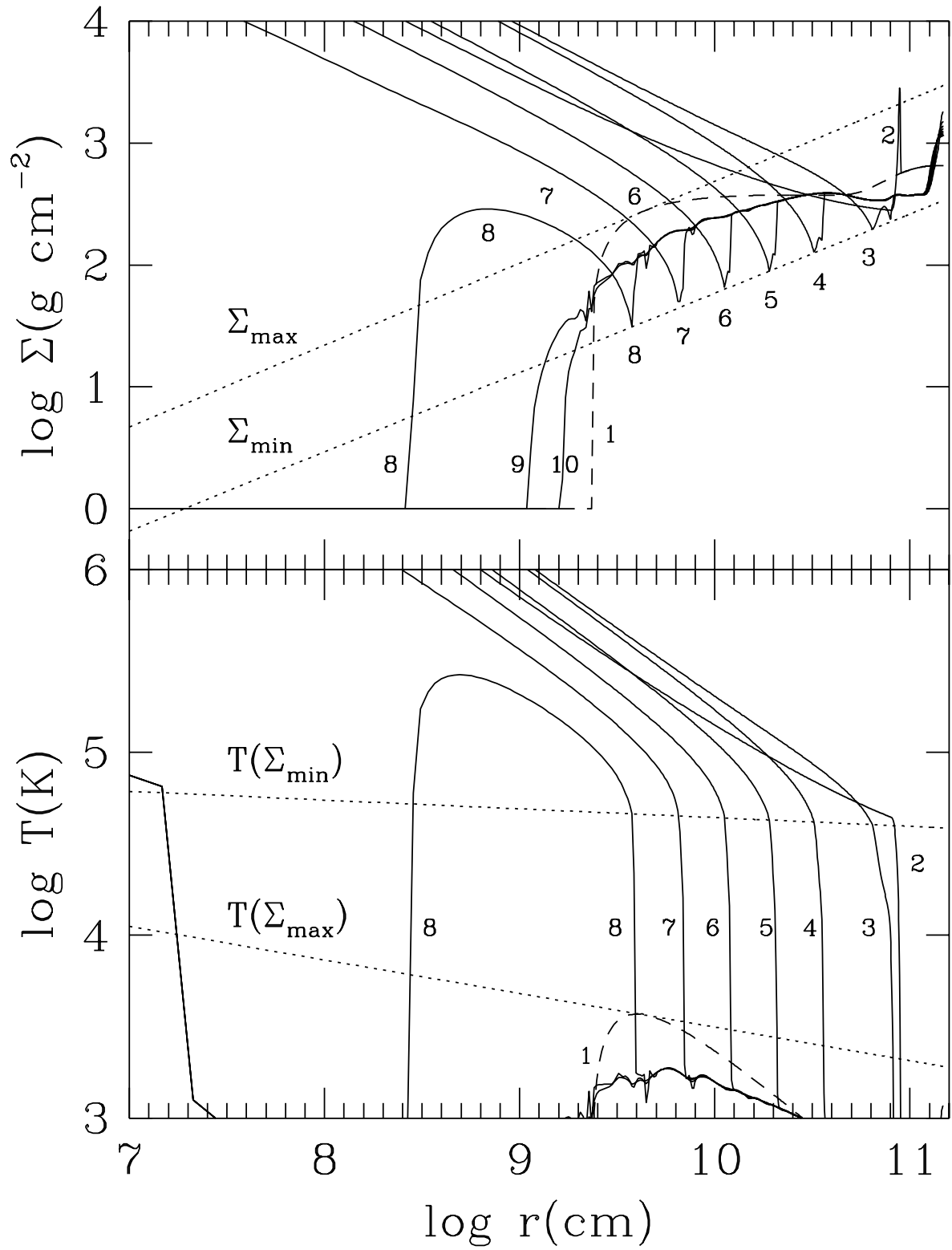
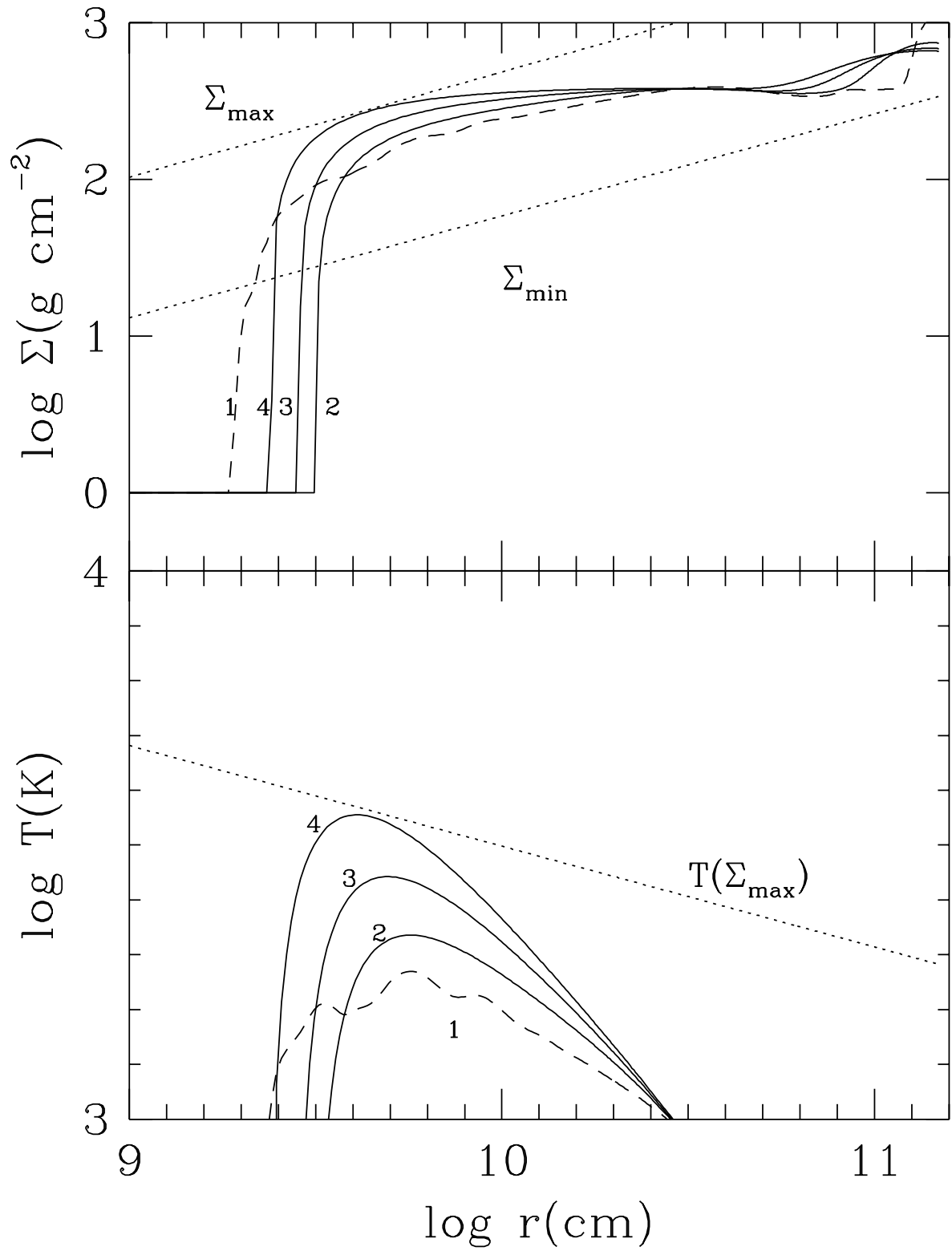
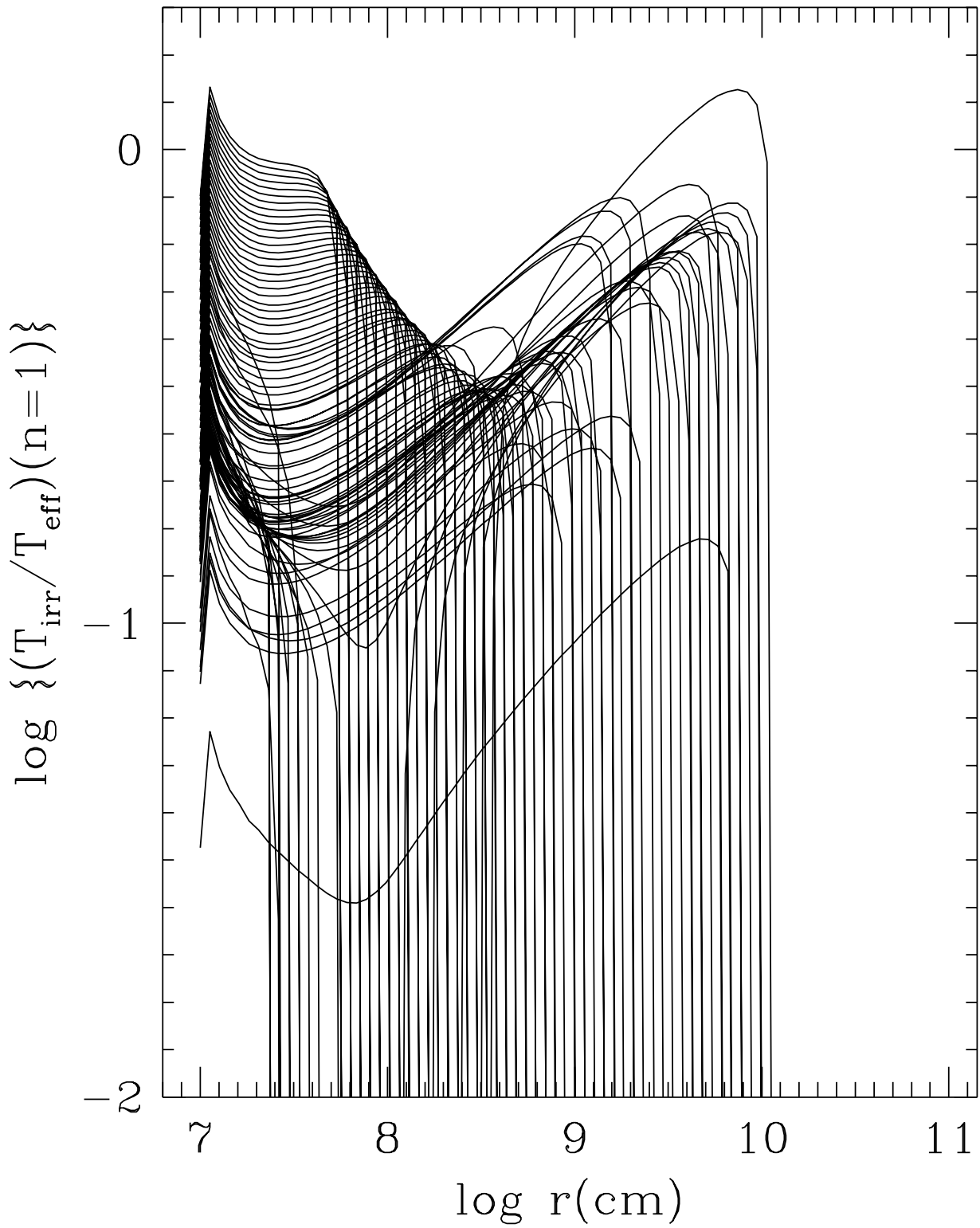
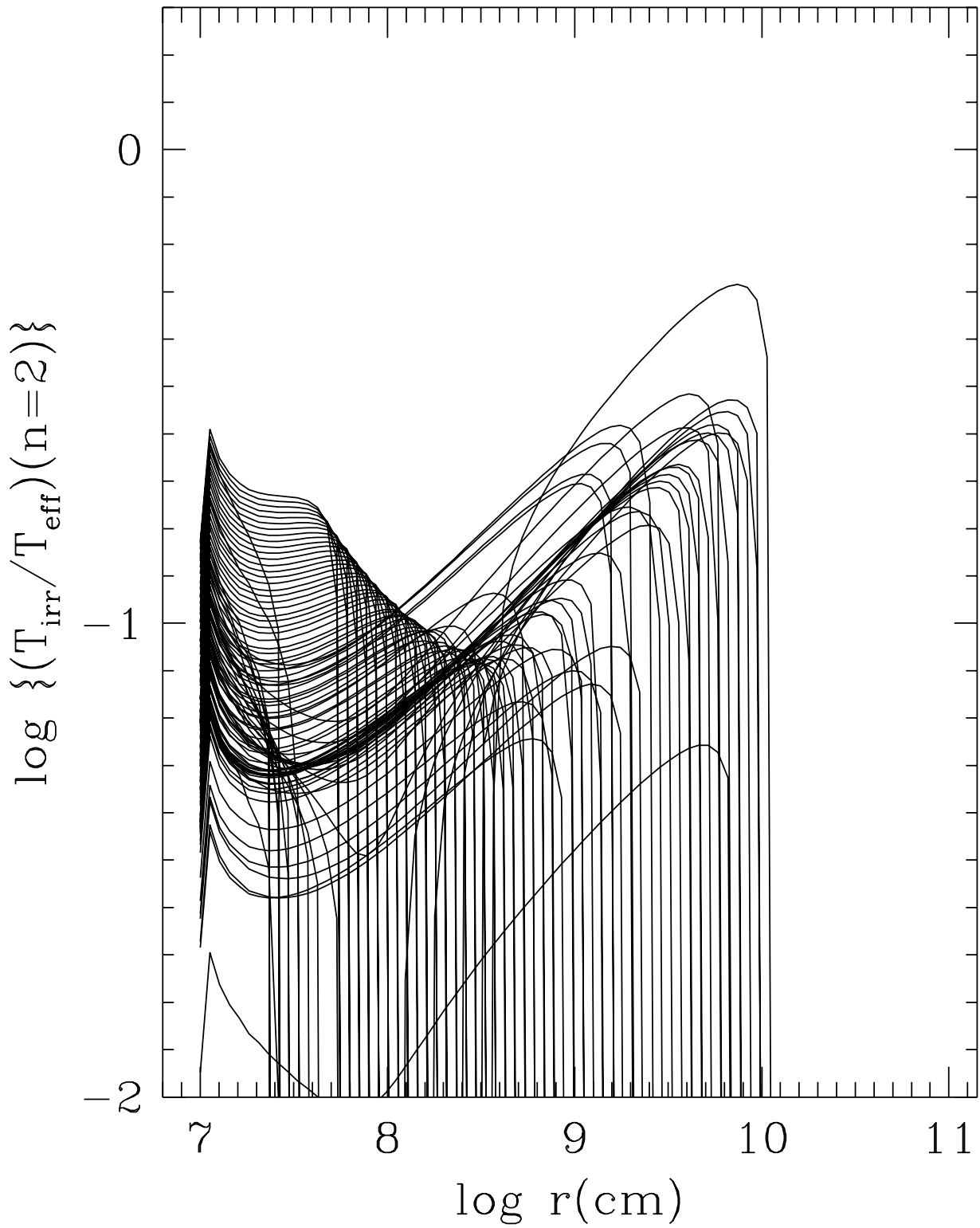
\end{document}